\documentclass[fleqn,10pt]{wlscirep}
\usepackage[utf8]{inputenc}
\usepackage[T1]{fontenc}
\usepackage{siunitx}
\usepackage{placeins}

\title{Laser spectroscopy of indium Rydberg atom bunches by electric field ionization}

\author[1,*]{A.~R.~Vernon}
\author[2]{C.~M.~Ricketts}
\author[2]{J.~Billowes}
\author[1]{T.~E.~Cocolios}
\author[2,8]{B.~S.~Cooper}
\author[2,8]{K.~T.~Flanagan}
\author[3,4]{R.~F.~Garcia~Ruiz}
\author[1]{F.~P.~Gustafsson}
\author[1,4]{G.~Neyens}
\author[2]{H.~A.~Perrett}
\author[5,+]{B.~K.~Sahoo}
\author[6]{Q.~Wang}
\author[7]{F.~J.~Waso}
\author[9]{X.~F.~Yang}

\affil[1]{KU Leuven, Instituut voor Kern- en Stralingsfysica, B-3001 Leuven, Belgium}
\affil[2]{School of Physics and Astronomy, The University of Manchester, Manchester M13 9PL, United Kingdom}
\affil[3]{EP Department, CERN, CH-1211 Geneva 23, Switzerland}
\affil[4]{Massachusetts Institute of Technology, Cambridge, MA 02139, USA}
\affil[5]{Atomic, Molecular and Optical Physics Division, Physical Research Laboratory, Navrangpura, Ahmedabad 380009, India}
\affil[6]{School of Nuclear Science and Technology, Lanzhou University, Lanzhou 730000, China}
\affil[7]{Stellenbosch University, Merensky Building, Merriman Street, Stellenbosch, South Africa}
\affil[8]{Photon Science Institute, Alan Turing Building, University of Manchester, Manchester M13 9PY, United Kingdom}
\affil[9]{School of Physics and State Key Laboratory of Nuclear Physics and Technology, Peking University, Beijing 100871, China}

\affil[*]{Corresponding author: adam.vernon@cern.ch}

\begin{abstract}
This work reports on the application of a novel electric field-ionization setup for high-resolution laser spectroscopy measurements on bunched fast atomic beams in a collinear geometry.
In combination with multi-step resonant excitation to Rydberg states using pulsed lasers, the field ionization technique demonstrates increased sensitivity for isotope separation and measurement of atomic parameters over non-resonant laser ionization methods.
The setup was tested at the Collinear Resonance Ionization Spectroscopy experiment at ISOLDE-CERN to perform high-resolution measurements of transitions in the indium atom from the 5s$^2$5d~$^2$D$_{5/2}$ and 5s$^2$5d~$^2$D$_{3/2}$ states to 5s$^2$($n$)p~$^2$P and 5s$^2$($n$)f~$^2$F Rydberg states, up to a principal quantum number of $n$~=~72.
The extracted Rydberg level energies were used to re-evaluate the ionization potential of the indium atom to be \SI{46670.1055(21)}{\per\centi\meter}.
The nuclear magnetic dipole and nuclear electric quadrupole hyperfine structure constants and level isotope shifts of the 5s$^2$5d~$^2$D$_{5/2}$ and 5s$^2$5d~$^2$D$_{3/2}$ states were determined for $^{113,115}$In.
The results are compared to calculations using relativistic coupled-cluster theory.
A good agreement is found with the ionization potential and isotope shifts, while disagreement of hyperfine structure constants indicates an increased importance of electron correlations in these excited atomic states.
With the aim of further increasing the detection sensitivity for measurements on exotic isotopes, a systematic study of the field-ionization arrangement implemented in the work was performed and an improved design was simulated and is presented.
The improved design offers increased background suppression independent of the distance from field ionization to ion detection.
\end{abstract}

\begin{document}

\flushbottom
\maketitle

\thispagestyle{empty}

\section*{Introduction}
The ability to separate and study small quantities of isotopes from a large ensemble without losses is the limiting factor of many experimental studies in modern nuclear physics \cite{Ruiz2019,Baumann2007a, DeGroote2020,Wienholtz2013}, as exotic isotopes of interest can often only be produced at low rates (fewer than 100s of ions per second) and their accumulation into substantial quantities is prevented by their short half-lives.
Furthermore, sensitive detection or separation of small quantities of isotopes also has numerous technological applications \cite{Fujiwara2019,Lynch2014,Lu2003,Swift2018,Walker1999,Kane1998a,Paisner1988}.
Fast beam collinear laser spectroscopy techniques have allowed high-precision measurements on short-lived isotopes, down to rates of fewer than 100 ions per second \cite{DeGroote2020,Miller2019b,GarciaRuiz2016}.
These approaches use the Doppler compression of an accelerated atomic beam to enable high-precision laser spectroscopy measurements to be performed in a collinear geometry \cite{Wing1976}.
This technique is now being implemented at radioactive ion beam facilities worldwide, giving a resolution of a few 10s of MHz, which is sufficient to resolve the hyperfine structure for nuclear physics studies \cite{Campbell2016,Neugart2017,Voss2016,Minamisono2013}.

Motivated by a need for higher sensitivity to access exotic isotopes produced at rates lower than a few ions per second, a variation of the technique, the Collinear Resonance Ionization Spectroscopy (CRIS) \cite{Flanagan2013, Kudriavtsev1982a} experiment at CERN-ISOLDE \cite{Catherall2017} has been developed.
The technique is based on resonant laser excitation of atom bunches \cite{Ricketts2019,Mane2009} using a high-resolution pulsed laser, followed by resonant or non-resonant ionization of the excited atoms.
The ions are then deflected away from the atoms which were not resonantly excited, allowing ion detection measurements with  significantly reduced background.
The experiment has so far reached a background suppression factor of \SI{2.5E8}{} providing a detection sensitivity down to yields of around 20 atoms per second \cite{DeGroote2017b}.
The main source of ion background for the technique is due to collisional re-ionization of the atom beam (often also containing a substantial amount of isobaric contamination) with residual gas atoms along the collinear laser overlap volume.
For this reason, considerable effort is given to reach ultra-high vacuum pressures ($<$\SI{1E-9}{mbar}) in this overlap region.
The work reported here demonstrates that the incorporation of field ionization, previously tested on continuous atom beams \cite{Dinger1986, Aseyev1993b}, can further increase the sensitivity of measurements on bunched atomic beams by also compressing the measurements into a narrow ionization volume.
In addition, we show the approach has advantages for measurements of atomic parameters when combined with the multi-step pulsed narrow-band laser excitation.
The sensitivity of the approach is further improved by removing the need for a powerful non-resonant laser ionization step, which often contributes substantially to the re-ionization background.
This is due to non-resonant ionization of contaminant atomic species, which are often neutralised into excited atomic levels\cite{Vernon2019}, that are easily ionized with powerful laser light.
Using the field-ionization technique, high-resolution measurements were performed of the energies of the 5s$^2$($n$)p $^2$P and 5s$^2$($n$)f $^2$F Rydberg series (intermittently over principal quantum numbers $n$~=~12 to $n$~=~72), which additionally enabled the re-evaluation of the ionization potential of the indium atom.
The production of atom bunches was implemented using an ablation ion source \cite{GarciaRuiz2018} with naturally abundant $^{113}$In and $^{115}$In isotopes.

In typical measurements using atomic transitions to extract nuclear structure parameters, both the upper and lower atomic states of the transition have a nuclear-structure dependence.
However due to the vanishing nuclear structure dependence of the Rydberg states \cite{Kopfermann1958, Niemax1980}, the transition measurements made here allowed extraction of the nuclear structure dependent parameters of the lower atomic states alone.
These measurements therefore allowed extraction of the hyperfine structure constants of the 5s$^2$5d $^2$D$_{5/2}$ and 5s$^2$5d $^2$D$_{3/2}$ states and their level isotope shifts (LISs), the shift in the level energies between the $^{113,115}$In isotopes. 
This allows direct comparison to atomic calculations of the isotope shift contribution from a single atomic state, in contrast to typical atomic transitions measurements where upper and lower state contributions are combined.
The experimentally determined LISs, hyperfine structure constants and indium atom ionization potential are compared to calculations from relativistic coupled-cluster theory \cite{Sahoo2018}.
Comparisons with direct measurements of atomic states of differing symmetry, such as $^2$D$_{5/2}$ and $^2$D$_{3/2}$, can give a valuable benchmark to assist the extraction of nuclear structure observables with higher accuracy \cite{Sahoo2020, GarciaRuiz2018}.
In addition, the development of accurate calculations for states of different symmetry has the potential to probe new aspects of the atomic nucleus and its interactions \cite{Berengut2018b,Flambaum2018a,Reinhard2019a, Allehabi2020, Flambaum2019a}.\\

\section*{Methods}
\subsection*{Experimental setup}
\begin{figure}[h]
\centering
\includegraphics[width=18cm]{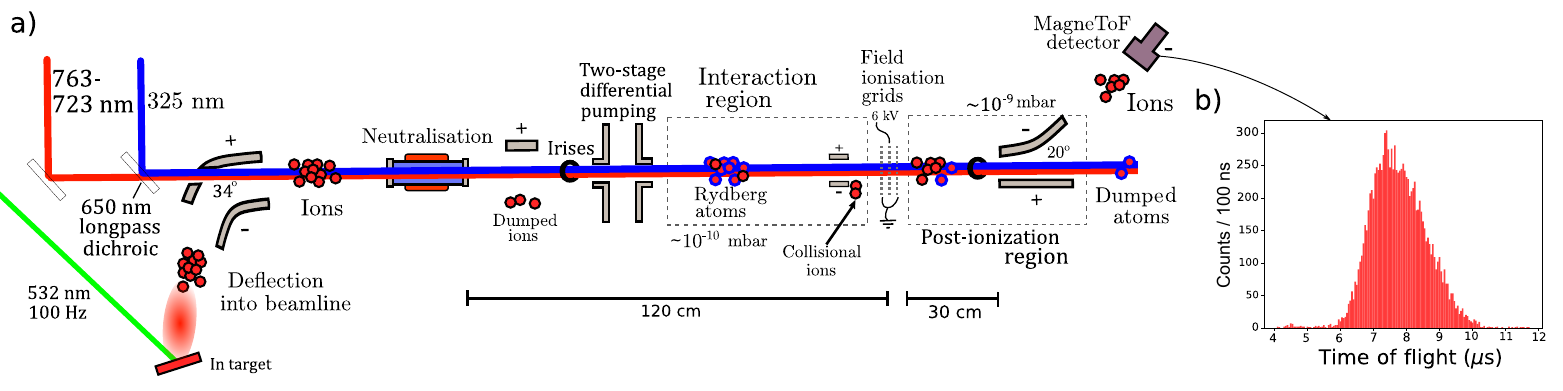}
\caption{a) A schematic diagram of the layout of the modified CRIS experiment, to use field ionization. Electrostatic deflectors guide the ions created by ablation to be neutralised. Atoms follow multi-step resonant laser excitation to Rydberg states, and are subsequently field ionized and counted by a particle detector.
b) The time-of-flight distribution of the bunch created by the ablation ion source.}
\label{fig:CRIS_layout_old}
\end{figure}

The measurements reported here were performed following a modification of the CRIS experimental setup \cite{Vernon2019a,Koszorus2019}.
A schematic layout of the modified setup is shown in Figure \ref{fig:CRIS_layout_old}a).
In$^+$ ions were produced using an ablation ion source (detailed in Refs. \cite{GarciaRuiz2018, Vernon2019a}), with a pulsed 532-nm Litron LPY 601 50-100 PIV Nd:YAG laser focused to produce a fluence of $>$0.5~J/cm$^2$ on a solid indium target (99\% purity).

This produced high-intensity bunches of indium ions at the 100~Hz repetition rate of the laser with a typical bunch width of \SI{2.0(3)}{\micro \second} (see Figure \ref{fig:CRIS_layout_old}b).
This ablation laser was used as the start trigger to synchronise the atomic bunches in the interaction region with the pulsed lasers subsequently used for spectroscopy.
A set of ion optics were used to focus and accelerate the In$^+$ ions to 25~keV and deflect them by 34$^\circ$ to overlap with two laser beams in a collinear geometry.
The acceleration creates a kinematic separation in the transition frequency of the naturally abundant isotopes $^{115}$In (95.72\%) and $^{113}$In (4.28\%), which greatly enhances the isotope selectivity of the approach compared to in-source laser spectroscopy separation or measurement techniques \cite{Fujiwara2019} such as laser induced breakdown spectroscopy (LIBS) \cite{Civis2014}, resonant ionization mass spectroscopy (RIMS) \cite{Wendt2000} or in-gas cell laser ionization spectroscopy (IGLIS) \cite{Zadvornaya2018}.

The ions were subsequently neutralised, using a sodium-filled charge-exchange cell heated to 300(10)$^\circ$C, with an efficiency of 60(10)\%, where 64\% of the atomic population is simulated to be in the 5s$^2$5p $^2$P$_{3/2}$ metastable state \cite{Vernon2019}. The ions which were not neutralised were deflected electrostatically following the cell.
After approximately 80~cm of flight (the maximum \SI{4}{\micro \second} bunch width corresponds to a spatial spread of 81~cm at 25~keV), the indium atoms were then excited using either the 5s$^2$5p~$^2$P$_{3/2}$ $\rightarrow$ 5s$^2$5d~$^2$D$_{5/2}$ (325.6~nm) or 5s$^2$5p~$^2$P$_{3/2}$ $\rightarrow$ 5s$^2$5d~$^2$D$_{3/2}$ (325.9~nm) transition, depending on the Rydberg series to be studied.\\

\begin{figure}[h]
\centering
\includegraphics[width=7cm]{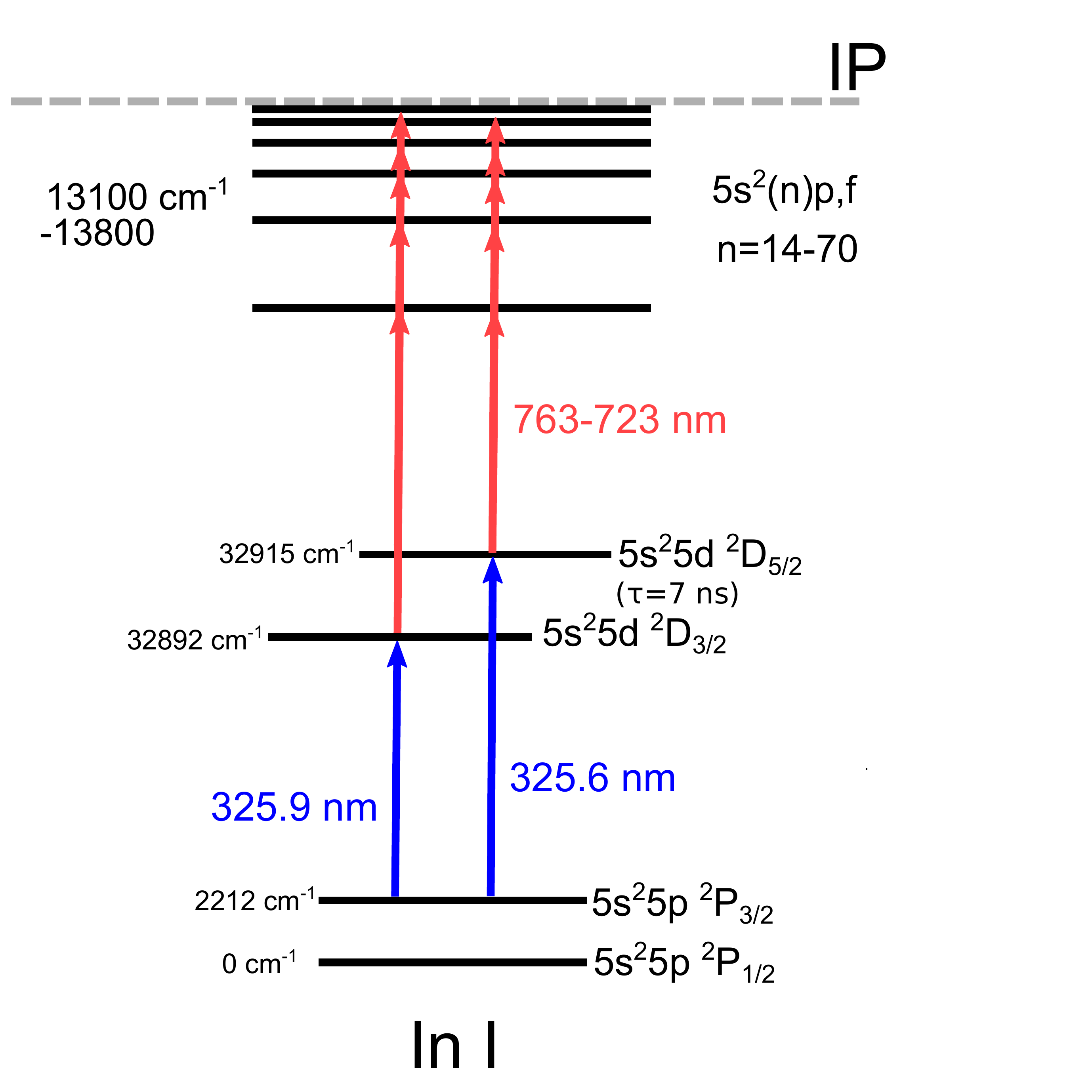}
\caption{The multi-step laser excitation schemes used in this work on the indium atom. Field ionization was used following the two laser steps to a Rydberg state. The lifetime of the intermediate 5s$^2$5d $^2$D$_{5/2}$, 5s$^2$5d $^2$D$_{3/2}$ states is 7.0(4)~ns \cite{Jonsson1983}.}
\label{fig:FI_scheme}
\end{figure}

The multi-step ionization schemes used in this work are shown in Figure \ref{fig:FI_scheme}.
The first step light was produced using a Spectron Spectrolase 4000 pulsed dye laser with DCM dye dissolved in ethanol, this produced fundamental light at 650~nm which was frequency doubled using a BaB$_2$O$_4$ crystal to 325~nm.
The linewidth of the laser was $\sim$14~GHz, allowing excitation of all of the hyperfine structure of the 5s$^2$5p $^2$P$_{3/2}^o$ state, while the 699~GHz separation between the 5s$^2$5d $^2$D$_{5/2}$ and 5s$^2$5d $^2$D$_{3/2}$ states required tuning the laser to each fine structure transition.
The dye laser was pumped with 532-nm light from the second output head of the Litron LPY 601 50-100 PIV Nd:YAG laser used for ablation. 
The second excitation step was scanned in laser frequency to perform high-resolution laser spectroscopy, from either the 5s$^2$5d $^2$D$_{5/2}$ or 5s$^2$5d $^2$D$_{3/2}$ state to a Rydberg state of the 5s$^2$($n$)f or 5s$^2$($n$)p series, ranging over $n$~=~14-72 (763-723~nm).
The high-resolution infrared laser light was produced using an injection-locked Ti:Sapphire laser \cite{Kessler2008, Sonnenschein2017} pumped by a LEE LDP-100MQ Nd:YAG laser and seeded using a narrowband continuous-wave Matisse Ti:Sapphire laser by Spectra-Physics.
This provided the pulsed narrowband (20(5)~MHz \cite{Sonnenschein2017}) laser light to be used for spectroscopy.
The resonantly excited indium Rydberg states were then field ionized in a longitudinal geometry by thin wire grids with a field gradient of \SI{7.5}{\kilo\volt\per\centi\meter}. Technical details of this setup are given in the \hyperlink{FIdesign}{Systematics of the field-ionization setup}~section.
Spatial alignment of the atom and ion paths was performed using irises and Faraday cups. The ion beam waist was measured to be around 3(1) mm using an iris \cite{Vernon2019a}, below this a reduction in beam current begin to be observed. This was measured $\sim$30 cm from the neutralisation cell.
Following ionization the ions were deflected by 20$^\circ$ onto a ETP DM291 MagneTOF\textsuperscript{TM} detector and the recorded count rate was used to produce the hyperfine spectra as a function of the infared laser frequency.

\subsection*{Coupled-cluster calculations}

In order to compare to our experimental results, the indium atom ionization potential, hyperfine structure constants A$_{\text{hf}}$ and B$_{\text{hf}}$, and atomic isotope shift factors were calculated using relativistic coupled-cluster (RCC) theory as outlined below.
The A$_{\text{hf}}$ and B$_{\text{hf}}$ constants were calculated using an expectation-value evaluation approach as described in Ref.~\cite{Sahoo2015}.
While the atomic parameters for the isotope shift were calculated using an analytic response approach (AR-RCC), an approach developed for increased accuracy for the evaluation of isotope shift contributions, as described in Ref.~\cite{Sahoo2020}.

In RCC theory, the wave function ($\vert \Psi_v \rangle$) of an atomic state with a closed-core and a valence orbital $v$ can be expressed as
\begin{eqnarray}
 \vert \Psi_v \rangle  = e^T [ 1+ S_v ] \vert \Phi_v \rangle ,
 \label{eqcc}
\end{eqnarray}
where $\vert \Phi_v \rangle$ is the mean-field wave function, defined as $\vert \Phi_v =a_v^{\dagger} \vert \Phi_0 \rangle$, with the Dirac-Hartree-Fock (DHF) wave function of the closed-core, $\vert \Phi_0 \rangle$.
Here, $T$ and $S_v$ are the RCC excitation operators which incorporate electron correlation effects by exciting electrons in $\vert \Phi_0 \rangle$ and $\vert \Phi_v \rangle$, respectively, to 
the virtual space.
The amplitudes of the RCC operators and energies were obtained by solving the following equations
\begin{eqnarray}
 \langle \Phi_K \vert \left (H e^T \right )_c   \vert \Phi_0 \rangle &=& \delta_{K,0} E_0 \; ,
\label{eqt}
 \end{eqnarray}
and
\begin{eqnarray}
 \langle \Phi_L \vert \left (H e^T \right )_c  S_v + \left (H e^T \right )_c \vert \Phi_v \rangle  &=&  \left ( \delta_{L,0} + \langle \Phi_L \vert S_v \vert \Phi_v \rangle \right ) E_v \; ,
\label{eqsv}
 \end{eqnarray}
where $H$ is the atomic Hamiltonian, and $\vert \Phi_K \rangle$ and $\vert \Phi_L \rangle$ denote the excited determinants with respect to $\vert \Phi_0 \rangle$ and $\vert \Phi_v \rangle$.
Here, $E_0$ and $E_v$ correspond to the energies of the closed-core and the closed-core with the valence orbital respectively.
Thus, the difference between $E_v$ and $E_0$ gives the binding energy or the negative of the ionization potential (IP) of the electron from the valence orbital, $v$.
The hyperfine structure constants of the unperturbed state were evaluated by
\begin{eqnarray}
\frac{\langle \Psi_v | O | \Psi_v \rangle} {\langle \Psi_v| \Psi_v \rangle}  
&=& \frac{\langle \Phi_v | [1+ S_v^{\dagger}] 
e^{T^{\dagger}} O e^T [1+S_v] | \Phi_v \rangle} {\langle \Phi_v| [1+ S_v{\dagger}] 
e^{T^{\dagger}} e^T [1+S_v] | \Phi_v \rangle } ,
\label{prpeq}
\end{eqnarray}
where $O$ is the hyperfine interaction operator. In the above expression, the non-terminating series of $e^{T^{\dagger}} O e^T$ and $e^{T^{\dagger}} e^T$ in the numerator and denominator, respectively, were calculated by adopting a self-consistent iterative procedure as described in Ref. \cite{Sahoo2015}.
All-possible singles and doubles excitations were included in the RCC calculations (RCCSD) for determining the energies and hyperfine structure constants.
The calculations were performed by first considering the Dirac-Coulomb (DC) Hamiltonian, then including the Breit and lower-order quantum electrodynamics (QED) interactions as described in Ref. \cite{Yu2019}.
Corrections due to the Bohr-Weisskopf (BW) effect to the hyperfine structure constants were estimated by considering a Fermi-charge distribution of nucleus.

The AR-RCC approach adopted to determine the field shift (FS), normal mass shift (NMS) and the specific mass shift (SMS) constants was implemented by evaluating the first order perturbed energies due to the respective operators, as discussed in Ref.~\cite{Sahoo2020}.
The AR-RCC theory calculations were also truncated using the singles and doubles excitation approximation (AR-RCCSD) when used to calculate the FS, NMS and SMS constants in this work.
Contributions from the DC Hamiltonian and corrections from the Breit and QED interactions were evaluated explicitly and are shown in Table~\ref{tab:is_consts}.

\section*{Analysis and results}

\begin{figure}[h]
\centering
\includegraphics[width=16cm]{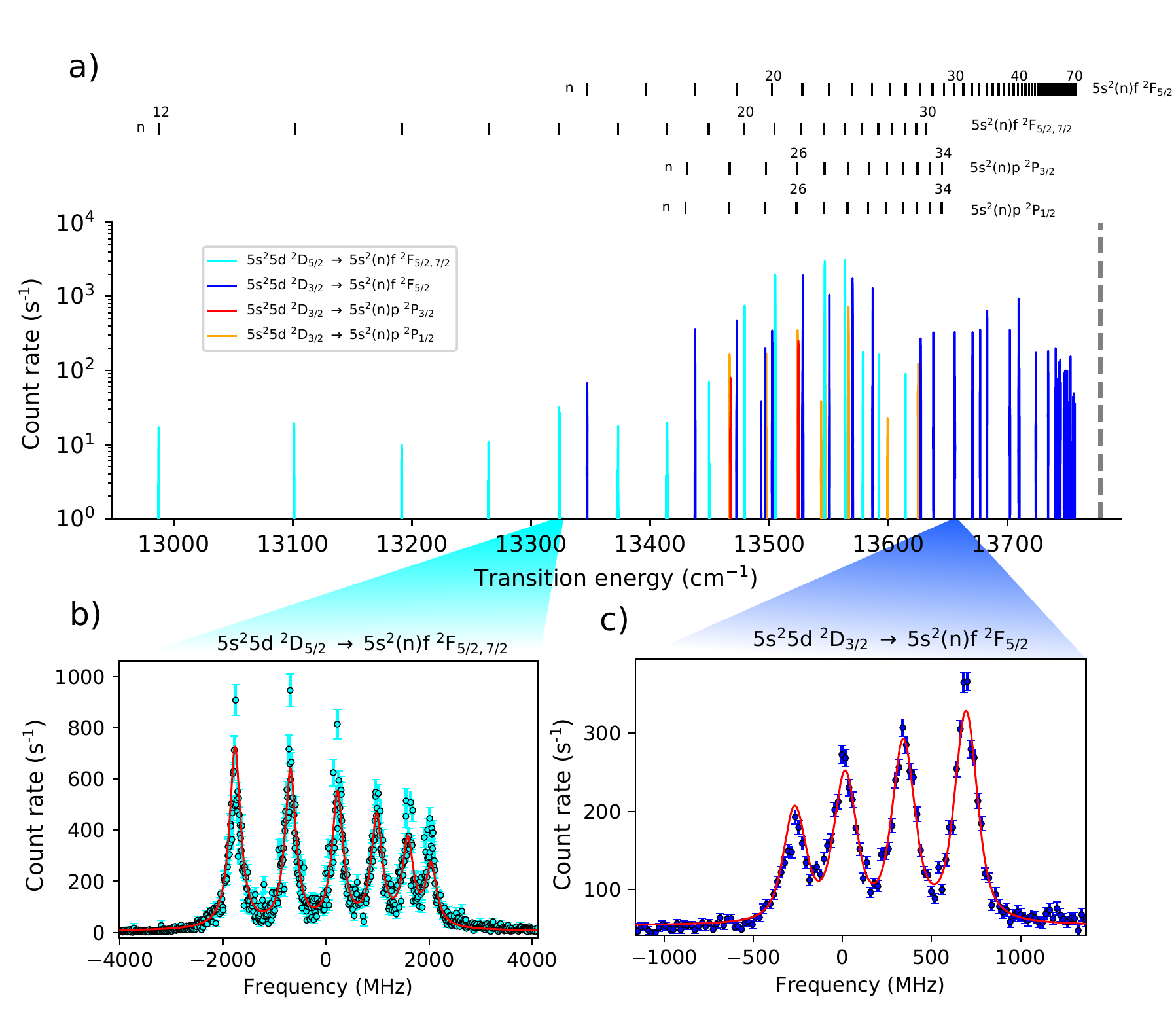}
\caption{Summary of the high-resolution measurements of the transitions to Rydberg states from the $^2$P$_{1/2}$ 5s$^2$($n$)p, $^2$P$_{3/2}$ 5s$^2$($n$)p and $^2$F$_{5/2}$ 5s$^2$($n$)f series, showing a) the spectrum of resonances measured in the 770-720~nm range, with an inset in black indicating energies for the members of the Rydberg series using Expression~\ref{E:n}. Example hyperfine spectra of $^{115}$In from the b) 5s$^2$5d $^2$D$_{5/2}$ and c) 5s$^2$5d $^2$D$_{3/2}$ lower states, resolvable from $n$~=~15-34. Fits to the spectra are indicated in red.}
\label{fig:Rydberg_summary}
\end{figure}

A summary of the high-resolution Rydberg series measurements is presented in Figure~\ref{fig:Rydberg_summary}. 
A range of wavelengths from 720 to 770~nm (14000-12900~cm$^{-1}$) were used to cover the transitions to Rydberg states in this work ($n$~=~12-72), as shown in Figure~\ref{fig:Rydberg_summary}a).
The energies of the states in the Rydberg series (vertical black dashes in Figure~\ref{fig:Rydberg_summary}a)) were estimated using the Rydberg formula \cite{Neijzen1982a} extrapolating from the energies of the five lowest principal quantum number atomic states ($n$~=~4-9 for ($n$)f, $n$~=~5-10 for ($n$)p), taken from literature \cite{NIST}.
See the Section~\hyperlink{eval}{Evaluation of the ionization potential of the indium atom} for details.
Figures \ref{fig:Rydberg_summary}b) and \ref{fig:Rydberg_summary}c) show hyperfine spectra obtained for the 5s$^2$5d $^2$D$_{5/2}$ $\rightarrow$ 5s$^2$($n$)f $^2$F$_{5/2, 7/2}$ and 5s$^2$5d $^2$D$_{3/2}$ $\rightarrow$  5s$^2$($n$)f $^2$F$_{5/2}$ transitions respectively.
The hyperfine structure resulting from the 5s$^2$5d $^2$D$_{5/2}$ and 5s$^2$5d $^2$D$_{3/2}$ states is visible in these spectra, while the contribution from the Rydberg state in both cases is vanishingly small due to the reduced overlap of the electronic wavefunctions at the nucleus \cite{Kopfermann1958, Niemax1980}.
The fine structure splitting between 5s$^2$($n$)f $^2$F$_{5/2}$ and 5s$^2$($n$)f $^2$F$_{7/2}$ Rydberg states has been measured to be $<$1~MHz \cite{Hong1995} and smaller than the linewidth of the laser used in this work.
The upper states of transitions from 5s$^2$5d $^2$D$_{5/2}$ are therefore denoted as 5s$^2$($n$)f $^2$F$_{5/2,7/2}$ to indicate that excitation to both the 5s$^2$($n$)f $^2$F$_{5/2}$ and 5s$^2$($n$)f $^2$F$_{7/2}$ states are included.
\\

\subsection*{Hyperfine structure constants and isotope shifts}

The extracted magnetic dipole and electric quadrupole hyperfine constants, A$_{\text{hf}}$ and B$_{\text{hf}}$, of the 5s$^2$5d $^2$D$_{5/2}$ and 5s$^2$5d $^2$D$_{3/2}$ states for $^{113,115}$In are displayed in Table \ref{tab:hfs_consts}.
The constants were determined by least-square minimisation fitting \cite{Newville2014} of the obtained hyperfine spectra to the well known hyperfine structure relations \cite{Schwartz1955} with A$_{\text{hf}}$ and B$_{\text{hf}}$ as free parameters.
A Voigt line profile \cite{Olivero1977} was used in the fitting with the Gaussian and Lorentzian components and transition intensities as free parameters.

The presented A$_{\text{hf}}$ and B$_{\text{hf}}$ values are an average of the results from well-resolved hyperfine spectra obtained in this work, over principal quantum numbers $n$~=~15-34.
For Rydberg states below $n\approx$20, the applied laser power saturated the transition and this led to power broadening which reduced the resolution of those hyperfine spectra.
The transitions above $n\approx$30 were not saturated (see Section \hyperlink{FIdesign}{Systematics of the field-ionization setup}), because the transition probability scales approximately\cite{Feneuille1982} as $n^{-3}$.
For these transitions a greater ablation laser fluence and ion source extraction potential were required to obtain a similar resonant signal level.
This resulted in an increased energy spread of the ion bunch, increasing the Gaussian contribution to the linewidth to greater than 100~MHz and obscuring the hyperfine spectra in those cases.
No statistically significant deviation was seen for contributions from the Rydberg states to the hyperfine structure constants down to $n$~=~12.
The larger uncertainty of the extracted B$_{\text{hf}}$ values was due to their small magnitudes compared to the laser linewidth of 20(5)~MHz.
While the larger uncertainty on the $^{113}$In values was due to the lower statistics, related to its lower natural abundance of 4.28\% \cite{Haynes}, in combination with a reduction in excitation efficiency due to the 14-GHz linewidth of the dye laser used for the first step transition which was centered at a frequency for $^{115}$In.
The isotope shifts for the levels (LIS) were also extracted and are displayed in Table \ref{tab:hfs_consts}.
In order to correct for the effect of an ion velocity distribution created in the ablation process on the measured LIS between $^{113,115}$In, the time of flights of $^{113,115}$In gated by the on-resonance laser frequencies were used to ensure LISs were measured from the same velocity component \cite{Vernon2019a, GarciaRuiz2018a} removing the Doppler shift from the presence of any distribution of velocities.\\

\begin{table}[h]
\centering
\begin{tabular}{|c c|c|c|c|c|c|}
\hline
 & &\multicolumn{2}{c}{$^{115}$In} & \multicolumn{2}{c}{$^{113}$In} &  \\
State & & A$_{\text{hf}}$ (MHz) & B$_{\text{hf}}$ (MHz) & A$_{\text{hf}}$ (MHz) & B $_{\text{hf}}$ (MHz) & LIS $^\ddagger$ (MHz)\\
\hline	
5s$^2$5d $^2$D$_{5/2}$ &\textbf{ Exp.} & \textbf{151.2(9)}& \textbf{33(20)} & \textbf{151(6)}  & \textbf{28(200)} & $-$\textbf{103(50)} \\
  & \textbf{Theor.  } &   \\
  & DHF      & 1.87  & 2.53 & 1.86  & 2.50 &  \\
  & RCCSD    & 39.66 & 24.84 & 39.57 & 24.53 & \\
  & $+$Breit & 39.80 & 24.84 & 39.71 & 24.53 & \\
  & $+$QED   & 40.00 & 24.99 & 39.91 & 24.68 & \\
  & $+$BW    & \textbf{39.87 }& \textbf{25.01} & \textbf{39.78 }& \textbf{24.70} & $-$\textbf{67.3(3.2)$^\dagger$}\\
  \hline
5s$^2$5d $^2$D$_{3/2}$ & \textbf{Exp.} & $-$\textbf{64(1)} & \textbf{41(20)} & $-$\textbf{68(5)} & \textbf{20(60)} & $-$\textbf{98(30)} \\
  & \textbf{Theor.}   &         &       &      &      &  \\
  & DHF      &  4.37   &  1.83 & 4.36 & 1.81 &  \\
  & RCCSD    & $-9.89$ & 17.82 & $-9.87$ & 17.60 &  \\
  & $+$Breit & $-9.89$ & 17.81 & $-9.87$ & 17.59 & \\
  & $+$QED   & $-9.87$ & 17.92 & $-9.85$ & 17.70 & \\
  & $+$BW    & $-$\textbf{9.74} & \textbf{17.98}& $-$\textbf{9.72} & \textbf{17.72} & $-$\textbf{67.7(3.2)}$^\dagger$ \\     
\hline
\end{tabular}
\caption{
\label{tab:hfs_consts} Hyperfine structure constants and LISs of the 5s$^2$5d $^2$D$_{5/2}$ and 5s$^2$5d $^2$D$_{3/2}$ states measured for $^{113,115}$In.\\
{\footnotesize $\dagger$ - Theoretical LISs were calculated using the F, K$_{\text{SMS}}$ and K$_{\text{NMS}}$ constants from Table~\ref{tab:is_consts} obtained from the AR-RCCSD approach, combined with the experimentally measured change in root-mean-square charge radius, $\delta \left\langle r^2 \right\rangle _{\mu}^{113, 115}$~=~-0.157(11)~fm$^2$, taken from Ref.~\cite{Fricke}, which gave the LIS uncertainty for the `calculated' values in the table shown in brackets.} \\
{\footnotesize $\ddagger$ - Here the sign convention of the $^{113}$In frequency centroid minus the $^{115}$In frequency centroid was used in the isotope shift formulae \cite{Sahoo2020}.}\\
}
\end{table}

The calculated A$_\text{hf}$ and B$_\text{hf}$ values of the 5s$^2$5d~$^2$D$_{3/2}$ and 5s$^2$5d~$^2$D$_{5/2}$ states using the RCCSD method are presented in Table~\ref{tab:hfs_consts}.
Literature nuclear magnetic dipole moment values of $\mu$~=~5.5289 $\mu_N$ (Ref.~\cite{Rice1957}) and nuclear electric quadrupole moment values of $Q$~=~0.80 b (Ref. \cite{GarciaRuiz2018}) for $^{113}$In, and of $\mu$~=~5.5408 $\mu_N$ (Ref. \cite{Flynn1960}) and $Q$~=~0.81 b (Ref. \cite{GarciaRuiz2018}) for $^{115}$In were used to evaluate the A$_\text{hf}$ and B$_\text{hf}$ values in Table~\ref{tab:hfs_consts}, from the calculated quantities of A$_\text{hf}I$/$\mu$ and B$_\text{hf}$/$Q$.
The DHF values of A$_{\text{hf}}$ were calculated to be 4.37 MHz and 1.87 MHz, whereas the RCCSD calculations gave $-$9.74~MHz and 39.87~MHz compared to the experimental values $-$64(2)~MHz and 151.2(9)~MHz for the 5s$^2$5d $^2$D$_{3/2}$ and 5s$^2$5d $^2$D$_{5/2}$ states of $^{115}$In, respectively.
The B$_\text{hf}$ values are within the 1$\sigma$ uncertainty of the experimental results, although the experimental uncertainty was large.
The difference in the BW correction to the A$_\text{hf}$ values between $^{113}$In and $^{115}$In were found to be negligibly small ($>$0.01~MHz).
Contributions from Breit and QED interactions were also found to be small.
This indicates that electron correlations due to core-polarization effects play the principal role in bringing the results close to the experimental values.
Thus, the experimental A$_{\text{hf}}$ values deviate in contrast to the LIS calculations using the AR-RCCSD calculations at the same level of truncation to singles and doubles excitation.
An explanation demands including triples excitations or employing a more rigorous theoretical approach for the evaluation of A$_{\text{hf}}$ factors.

In Table~\ref{tab:hfs_consts}, a comparison is also made between calculated LIS values with the measurements for the 5s$^2$5d $^2$D$_{3/2}$ and 5s$^2$5d $^2$D$_{5/2}$ states. 
The calculated FS (F), NMS (K$_\text{NMS}$) and SMS (K$_\text{SMS}$) constants, used to determine the calculated LIS values, are reported in Table~\ref{tab:is_consts} along with the included corrections.
For comparison to our relativistic {\it ab-initio} calculations of K$_\text{NMS}$, the K$_\text{NMS}$ constant values from the non-relativistic approximation are also shown, estimated by the relation K$_\text{NMS}=E_i m_e$ and experimental energies \cite{NIST}, $E_i$.
Unlike the A$_\text{hf}$ hyperfine structure constants, we find a good agreement between the measured and theoretical values for the LISs of the 5s$^2$5d $^2$D$_{3/2}$ and 5s$^2$5d $^2$D$_{5/2}$ states by substituting the calculated IS constants.\\

\begin{table}[h]
\centering
\begin{tabular}{|ccccc|}
\hline
 State & Method & F (MHz/fm$^2$) & K$_{\text{NMS}}$ (GHz amu) & K$_{\text{SMS}}$ (GHz amu) \\
\hline
5s$^2$5d $^2$D$_{5/2}$ & DHF & $\sim 0.0$ & 277.95 & $-38.48$ \\
  & \textbf{AR-RCCSD} & 293.33 & 198.76 & $-55.92$ \\
  & $+$Breit & 293.56 & 198.72 & $-55.41$ \\
  & $+$QED   & 288.39 & \textbf{198.97} & $-56.41$ \\
  & \textbf{Exp.}     &     & \textbf{226.20777(13)} &        \\ 
  \hline
5s$^2$5d $^2$D$_{3/2}$ & DHF & $\sim 0.0$ & 277.78 & $-37.39$\\
  & \textbf{AR-RCCSD} & 295.99 & 198.59 & $-55.69$ \\
  & $+$Breit & 296.19 & 198.54 & $-55.22$ \\
  & $+$QED   & 291.00 & \textbf{198.88 }& $-56.28$ \\
   & \textbf{Exp.}     &    & \textbf{226.59111(13) }&        \\ 
\hline
\end{tabular}
\caption{\label{tab:is_consts} Calculated F, K$_{\text{SMS}}$ and K$_{\text{NMS}}$ constants of the 5s$^2$5d $^2$D$_{5/2}$ and 5s$^2$5d $^2$D$_{3/2}$ states for the In atom using DHF and AR-RCCSD approaches.
Experimental level energies for the non-relativistic K$_\text{NMS}=E_i m_e$ approximation were taken from Ref.~\cite{NIST}. 
The factors were used to calculate the LIS values given in Table~\ref{tab:hfs_consts}, using the expression LIS~=~F$\delta \left\langle r^2 \right\rangle _{\mu}^{113, 115}$ + $\mu$(K$_{\text{NMS}}$ + K$_{\text{SMS}}$), where $\mu$~=~(m$_{113}$+m$_{115}$)/(m$_{113}$m$_{115}$) is the mass modification factor using atomic masses from Ref.~\cite{Wang2017}
}
\end{table}
\FloatBarrier

\subsection*{Rydberg state energies}

In order to reduce the systematic error of the measured transition frequencies, reference scans were performed every few hours using transitions to the 5s$^2$21f $^2$F$_{5/2}$ state. 
When the 5s$^2$5p~$^2$P$_{3/2}$ $\rightarrow$ 5s$^2$5d~$^2$D$_{3/2}$ first step transition was used, this was performed using the 5s$^2$5d $^2$D$_{3/2}$ $\rightarrow$ 5s$^2$21f $^2$F$_{5/2}$ transition.
While for the 5s$^2$5p~$^2$P$_{3/2}$ $\rightarrow$ 5s$^2$5d~$^2$D$_{5/2}$ first step transition, the 5s$^2$5d $^2$D$_{5/2}$ $\rightarrow$ 5s$^2$21f $^2$F$_{5/2}$ transition was used.
This allowed measurements of the Rydberg series to be referenced to the same 5s$^2$21f $^2$F$_{5/2}$ state.
The absolute energy of the 5s$^2$21f $^2$F$_{5/2}$ state was determined for the first time in this work, using an average of measurements of the 5s$^2$5d $^2$D$_{5/2}$ $\rightarrow$ 5s$^2$21f $^2$F$_{5/2, 7/2}$ and 5s$^2$5d $^2$D$_{3/2}$ $\rightarrow$  5s$^2$21f $^2$F$_{5/2}$ transition energies, combined with literature values for the 5s$^2$5d $^2$D$_{5/2}$ and 5s$^2$5d $^2$D$_{3/2}$ states, taken from Ref. \cite{George1990b}.
This gave an averaged value of 46420.309(5)~cm$^{-1}$ (1391645845(138)~MHz) for the energy of the 5s$^2$21f $^2$F$_{5/2}$ state, as presented in Table~\ref{tab:n21ref}.
This was the largest contribution to the final energy level uncertainty.
Other sources of systematic uncertainty to the absolute energy measured for the 5s$^2$21f $^2$F$_{5/2}$ state are also presented in Table~\ref{tab:n21ref}.
Where $\sigma(\text{T}_\text{B})$ is the systematic ion source voltage uncertainty, which determines the uncertainty in the energy of the atom beam, $\text{T}_\text{B}$, and therefore the transition energy through the Doppler shift. 
And $\sigma(\lambda)$ is the manufacture quoted absolute accuracy of the HighFinesse WSU2 wavemeter used. The wavemeter was drift stabilized by simultaneous measurement of a Toptica DLC DL PRO 780 diode laser locked to the 5s$^2$S$_{1/2}$~$\rightarrow$~5p $^2$P$_{3/2}$ F~=~2~-~3 transition of $^{87}$Rb using a saturated absorption spectroscopy unit \cite{Koszorus2019a}.
Transitions to the other principal numbers of the 5s$^2$($n$)f and 5s$^2$($n$)p series were then correlated with the closest 5s$^2$21f $^2$F$_{5/2}$ reference scans in time to determine the relative centroid shift of their hyperfine structure.
These centroid shifts are presented in Tables \ref{tab:fromd52} and \ref{tab:fromd32}  for the series measured using the  5s$^2$5p $^2$P$_{3/2}$ $\rightarrow$ 5s$^2$5d $^2$D$_{5/2}$ (325.6~nm) or 5s$^2$5p $^2$P$_{3/2}$ $\rightarrow$ 5s$^2$5d $^2$D$_{3/2}$ (325.9~nm) as first step transitions respectively.
The centroid shifts for the 5s$^2$($n$)f and 5s$^2$($n$)p series were then used to determine their absolute energy levels using the absolute value for the 5s$^2$21f $^2$F$_{5/2}$ state, as reported in Table~\ref{tab:n21ref}.\\

\begin{table}[h]
\centering
\begin{tabular}{|m{4.5cm}|c|c|c|c|}
\hline
 Transition & 5s$^2$21f $^2$F$_{5/2}$ (MHz) & $\sigma$ Lit. \cite{George1990b}  (MHz) & $\sigma(\text{T}_\text{B})$ (MHz) & $\sigma(\lambda)$  (MHz)\\
\hline
5s$^2$5d $^2$D$_{5/2}$ $\rightarrow$ 5s$^2$21f $^2$F$_{5/2,7/2}$ & 1391645920(166) & 150 & 70 & 2 \\
5s$^2$5d $^2$D$_{3/2}$ $\rightarrow$ 5s$^2$21f $^2$F$_{5/2}$& 1391645782(151) & 150 & 19 & 2 \\
\cmidrule{1-2}
Average & \textbf{1391645845(138)} &  &  & \\
\hline
\end{tabular}
\caption{\label{tab:n21ref} Determination of the absolute energy level of the 5s$^2$21f $^2$F$_{5/2}$ reference state. The final value is the weighted mean from two sets of reference transition measurements, 5s$^2$5d $^2$D$_{5/2}$ $\rightarrow$ 5s$^2$21f $^2$F$_{5/2,7/2}$ and 5s$^2$5d $^2$D$_{3/2}$ $\rightarrow$ 5s$^2$21f $^2$F$_{5/2}$.\\
`Lit.' refers to the uncertainty on the lower state energy taken from literature \cite{George1990b}.}
\end{table}

The energy levels for the members of the 5s$^2$($n$)f $^2$F$_{5/2}$ and 5s$^2$($n$)f $^2$F$_{5/2, 7/2}$ series shown in Tables~\ref{tab:fromd32} and \ref{tab:fromd52}, and Figure~\ref{fig:FI_scheme} have agreement between them, using the evaluated energy of the 5s$^2$21f$^2$F$_{5/2}$ reference from Table~\ref{tab:n21ref}.
The few principal quantum numbers with available values in literature \cite{Neijzen1981}, for the 5s$^2$($n$)f $^2$F$_{5/2, 7/2}$, 5s$^2$($n$)p $^2$P$_{1/2}$ and 5s$^2$($n$)p $^2$P$_{3/2}$ states, have agreement well within uncertainty.\\

\begin{table}[h]
\centering
\footnotesize
\begin{tabular}{|c|c|c|c|c|c|}
\hline
     &      &  &      & Literature &    \\
     &      & Centroid shift &  Energy level & energy level\cite{NIST} &    \\
Series & $n$ &  (MHz)      &     (cm$^{-1}$)   & (cm$^{-1}$)  &  $\delta_n$  \\
\hline
$^2$F$_{5/2,7/2}$  	&	12	&	  -15510479(20) 	&	  45902.9328(6)[46] 	& 45902.92(22)&	  0.04010(4) \\
5s$^2$($n$)f	&	13	&	  -12098644(10) 	&	  46016.7393(5)[46] 	&	&	  0.04028(5) \\
     	&	14	&	   -9393432(60) 	&	    46106.976(2)[5] 	&	&	  0.04052(8) \\
     	&	15	&	   -7212037(60) 	&	    46179.739(2)[5] 	&	&	   0.0407(1) \\
     	&	16	&	   -5427634(80) 	&	    46239.260(3)[5] 	&	&	   0.0408(1) \\
     	&	18	&	   -2710813(20) 	&	  46329.8837(7)[46] 	&	&	   0.0406(1) \\
     	&	20	&	     -769186(6) 	&	  46394.6494(2)[46] 	&	&	   0.0407(2) \\
     	&	21	&	0	&	  46420.3070(5)[46] 	&	&	   0.0408(2) \\
     	&	22	&	     666548(10) 	&	  46442.5403(3)[46] 	&	&	   0.0408(2) \\
     	&	23	&	   1247917(400) 	&	        46461.93(1) 	&	&	   0.0408(9) \\
     	&	24	&	   1758074(100) 	&	    46478.950(5)[5] 	&	&	   0.0407(6) \\
     	&	25	&	    2208136(60) 	&	    46493.962(2)[5] 	&	&	   0.0406(5) \\
     	&	26	&	    2607211(20) 	&	  46507.2739(7)[46] 	&	&	   0.0404(4) \\
     	&	28	&	    3280668(20) 	&	  46529.7380(7)[46] 	&	&	   0.0403(5) \\

\hline
\end{tabular}
\caption{\label{tab:fromd52} Energy levels of the $^2$F$_{5/2, 7/2}$ 5s$^2$($n$)f Rydberg series states determined from energy level shifts relative to the $^2$F$_{5/2, 7/2}$ 5s$^2$21f reference state, alongside the quantum defects, $\delta_n$, of the levels. Statistical uncertainty is indicated in parenthesis. Systematic uncertainty from the reference state is indicated in braces.}
\end{table}

\begin{table}[h]
\footnotesize
\centering
\begin{tabular}{|c|c|c|c|c|c|}
\hline
     &      &  &      & Literature &    \\
     &      & Centroid shift &  Energy level & energy level \cite{Neijzen1982a} &    \\
Series & $n$ &  (MHz)      &     (cm$^{-1}$)   & (cm$^{-1}$)  &  $\delta_n$  \\
\hline
$^2$P$_{1/2}$ & 22 & -1842830(20) &  46358.8365(7)[46] &  & 3.2238(2) \\
5s$^2$($n$)p & 23 &    -922829(20) &  46389.5244(6)[46] & &  3.2236(2) \\
     & 24 &    -132449(20) &  46415.8886(6)[46] & 46415.871(26) &  3.2235(2) \\
     & 25 &      551531(5) &  46438.7037(2)[46] & & 3.2233(2) \\
     & 26 &     1147394(6) &  46458.5796(2)[46] &  & 3.2231(2) \\
     & 28 &    2129967(60) &    46491.355(2)[5] & 46491.338(26) &   3.2229(2) \\
     & 30 &    2900676(60) &    46517.063(2)[5] & 46517.043(26) &  3.2226(2) \\
     \hline
$^2$P$_{3/2}$ & 22 &   -1816214(20) &  46359.7243(6)[46] &  & 3.1969(2) \\
5s$^2$($n$)p & 24 &     -112789(7) &  46416.5444(2)[46] & 46416.526(26) & 3.1966(2) \\
     \hline
$^2$F$_{5/2}$  	&	16	&	   -5427114(30) 	&	   46239.278(1)[46] 	&	&	  0.0404(1) \\
5s$^2$($n$)f	&	18	&	   -2711018(10) 	&	  46329.8768(4)[46] 	&	&	  0.0408(1) \\
     	&	19	&	    -1663438(5) 	&	  46364.8203(2)[46] 	&	&	  0.0409(1) \\
     	&	20	&	    -769307(10) 	&	  46394.6453(4)[46] 	&	&	  0.0409(2) \\
     	&	21	&	0	&	  46420.3070(5)[46] 	&	&	  0.0408(2) \\
     	&	22	&	      666346(5) 	&	  46442.5336(2)[46] 	&	&	  0.0411(2) \\
     	&	23	&	    1247803(10) 	&	  46461.9289(5)[46] 	&	&	  0.0410(3) \\
     	&	24	&	     1757899(8) 	&	  46478.9439(3)[46] 	&	&	  0.0410(3) \\
     	&	25	&	   2208075(100) 	&	    46493.960(5)[5] 	&	&	  0.0407(7) \\
     	&	26	&	   2607218(100) 	&	    46507.274(5)[5] 	&	&	  0.0404(7) \\
     	&	27	&	    2962711(20) 	&	  46519.1321(7)[46] 	&	&	  0.0402(5) \\
     	&	28	&	  3280581(4000) 	&	         46529.7(1) 	&	&	    0.04(1) \\
     	&	30	&	     3823828(9) 	&	  46547.8558(3)[46] 	&	&	  0.0401(6) \\
     	&	32	&	    4268023(10) 	&	  46562.6726(4)[46] 	&	&	  0.0410(7) \\
     	&	33	&	    4460517(30) 	&	    46569.093(1)[5] 	&	&	  0.0409(9) \\
     	&	34	&	    4636375(60) 	&	    46574.959(2)[5] 	&	&	   0.040(1) \\
     	&	38	&	    5205941(60) 	&	    46593.958(2)[5] 	&	&	   0.040(2) \\
     	&	40	&	     5428856(6) 	&	  46601.3938(2)[46] 	&	&	   0.039(1) \\
     	&	45	&	    5861534(20) 	&	  46615.8264(8)[46] 	&	&	   0.039(2) \\
     	&	50	&	    6170831(20) 	&	  46626.1434(6)[46] 	&	&	   0.042(3) \\
     	&	53	&	   6316157(200) 	&	    46630.991(5)[5] 	&	&	   0.037(7) \\
     	&	54	&	   6359037(200) 	&	    46632.421(7)[5] 	&	&	   0.042(8) \\
     	&	55	&	   6399878(300) 	&	        46633.78(1) 	&	&	    0.04(1) \\
     	&	57	&	    6475198(50) 	&	    46636.296(2)[5] 	&	&	   0.034(5) \\
     	&	60	&	   6574308(600) 	&	        46639.60(2) 	&	&	    0.03(2) \\
     	&	61	&	    6603688(70) 	&	    46640.582(2)[5] 	&	&	   0.040(7) \\
     	&	62	&	    6631893(20) 	&	  46641.5228(7)[46] 	&	&	   0.046(6) \\
     	&	65	&	    6709179(30) 	&	    46644.101(1)[5] 	&	&	   0.048(7) \\
     	&	68	&	   6776687(200) 	&	    46646.353(5)[5] 	&	&	    0.04(1) \\
     	&	69	&	    6796960(60) 	&	    46647.029(2)[5] 	&	&	    0.05(1) \\
     	&	70	&	   6816857(100) 	&	    46647.693(4)[5] 	&	&	    0.04(1) \\
     	&	72	&	   6853899(300) 	&	    46648.928(9)[5] 	&	&	    0.03(2) \\
\hline
\end{tabular}
\caption{\label{tab:fromd32} Energy levels of the $^2$P$_{1/2}$ 5s$^2$($n$)p, $^2$P$_{3/2}$ 5s$^2$($n$)p and $^2$F$_{5/2}$ 5s$^2$($n$)f Rydberg series states determined from energy level shifts relative to the $^2$F$_{5/2}$ 5s$^2$21f reference state, alongside the quantum defects, $\delta_n$, of the levels. Statistical uncertainty is indicated in parenthesis. Systematic uncertainty from the reference state is indicated in braces.}
\end{table}

\FloatBarrier

\subsection*{\hypertarget{eval}{Evaluation of the ionization potential of the indium atom}}
\label{section:IP}

\begin{figure}[h]
\centering
\includegraphics[width=17cm]{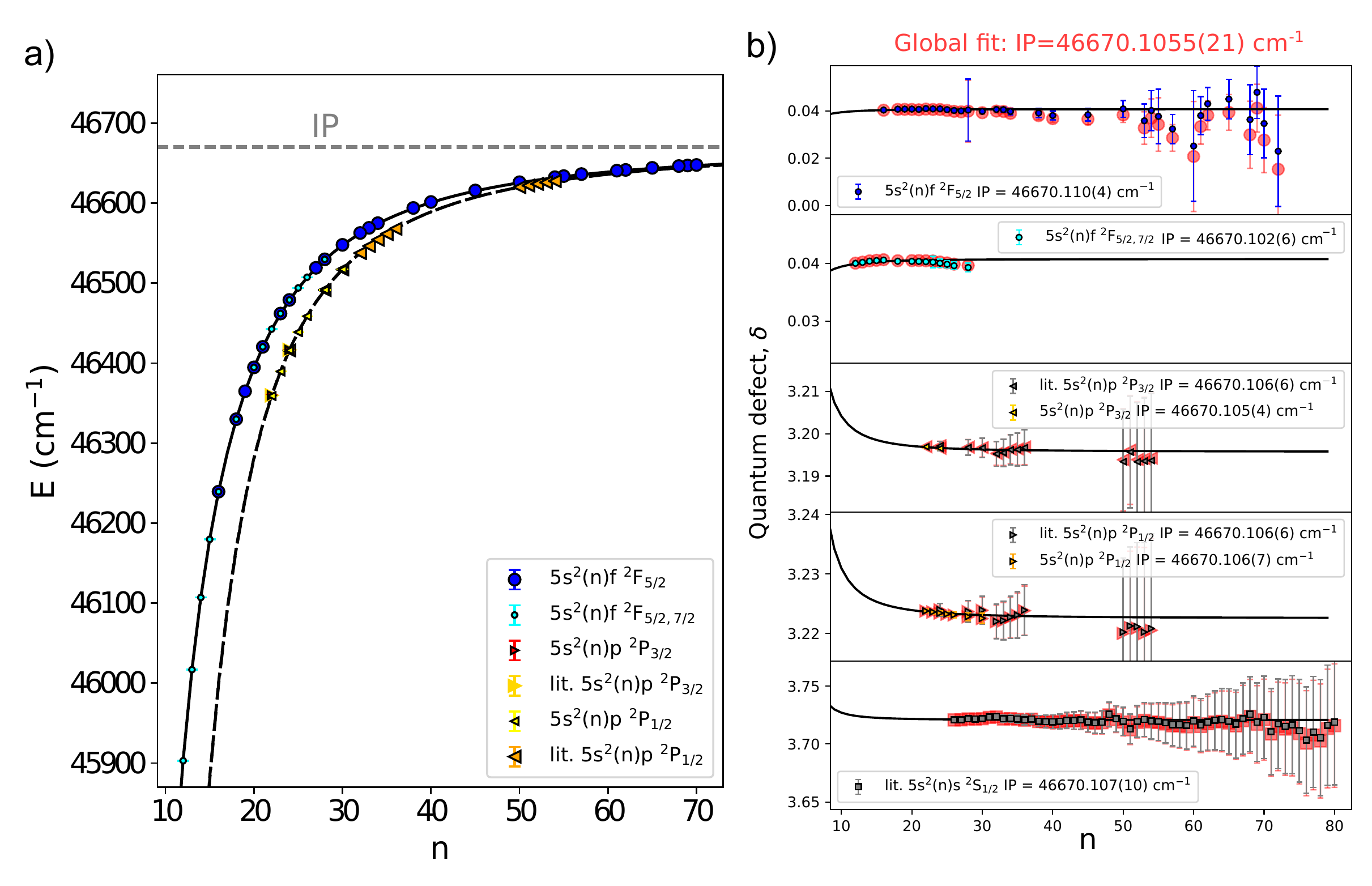}
\caption{\label{fig:simul_IP}
a) Extracted energies of the 5s$^2$($n$)p~$^2$P$_{1/2}$, 5s$^2$($n$)p~$^2$P$_{3/2}$, 5s$^2$($n$)f~$^2$F$_{5/2}$ and 5s$^2$($n$)f~$^2$F$_{5/2, 7/2}$ Rydberg series measurements. 
Literature values, labeled as `lit.', were taken from Ref.~\cite{Neijzen1982a}.
The black lines indicates the Rydberg expression values for the $^2$F$_{5/2, 7/2}$~5s$^2$($n$)f (solid line) and $^2$P$_{1/2, 3/2}$~5s$^2$($n$)p series (dashed line).
The ionization potential is indicated by the dashed grey line.
b) Determination of the ionization potential of the indium atom using a global simultaneous fit using Expressions \ref{E:n} and \ref{rydberg:ritz} to $\delta_n$ values of the 5s$^2$($n$)f $^2$F$_{5/2}$, $^2$P$_{1/2,3/2}$ 5s$^2$($n$)p, $^2$P$_{1/2,3/2}$ 5s$^2$($n$)p and $^2$P$_{1/2,3/2}$ 5s$^2$($n$)p of this work, along with literate (`lit') values for the $^2$P$_{1/2,3/2}$ 5s$^2$($n$)p series \cite{Neijzen1982a} and $^2$S$_{1/2}$ 5s$^2$($n$)s \cite{Neijzen1981a} series.
The red markers indicate series $\delta_n$ values after the global fitting (IP = 46670.1055(21)~cm$^{-1}$).
The IP value determined from the fit to each series independently are shown in their corresponding subplots.
The black lines indicate series $\delta_n$ values determined by Expression \ref{rydberg:ritz} from lowest-lying states \cite{NIST} ($n$~=~4-10 for ($n$)f, $n$~=~5-10 for ($n$)p and $n$~=~6-10 for ($n$)s)}
\end{figure}

The energy levels of the $^2$P$_{1/2}$~5s$^2$($n$)p, $^2$P$_{3/2}$~5s$^2$($n$)p, $^2$F$_{5/2}$~5s$^2$($n$)f and $^2$F$_{5/2, 7/2}$~5s$^2$($n$)f Rydberg series states determined in this work are shown in Figure \ref{fig:FI_scheme} in comparison to the accepted literature ionization potential (IP) of the indium atom \cite{Haynes}.
The energies of the the $n^{\text{th}}$ Rydberg series states, E$_n$, can be determined using the Rydberg expression \cite{Rothe2013, Foot2004}
\begin{equation}
\label{E:n}
    E_n = \text{IP} + \frac{R_{\text{115In}}}{(n - \delta_n)^2} \; ,
\end{equation}
where $\delta_n$ is the quantum defect \cite{Drake1994}, a measure of the difference in electronic structure for the Rydberg series of a multi-electron atom compared to hydrogen, included as the effective principal quantum number $n^* =  $n$ - \delta_n$.
The effect due to the finite mass of $^{115}$In compared to the electron, is given by the Rydberg constant $R_{\text{115In}}$ \cite{Foot2004} of \SI{109736.79}{\per\centi\meter}, which was derived from Penning trap atomic nuclei mass measurements \cite{Mount2009,Wieslander2009}.
Expression \ref{E:n} can be fitted to the experimental energy levels, leaving the IP and $\delta_n$ as free parameters. The result of this is shown by the black lines in Figure \ref{fig:simul_IP}a). Expression \ref{E:n} was fitted to lower-lying $n$ states of the series \cite{NIST} to predict laser frequency scan ranges and give $n$ assignments.

The $\delta_n$ values evaluated from our experimental energies are shown in Figure~\ref{fig:simul_IP}b)   alongside the Ritz expansion values (black lines) from \cite{Martin1980}
\begin{equation}
    \delta_n = a_0 + \frac{a_1}{(n-\delta_0)^2} + \frac{a_2}{(n-\delta_0)^4} + \frac{a_3}{(n-\delta_0)^6} \; \; ,
    \label{rydberg:ritz}
\end{equation}
where $a_{0,1,2,3}$ and $\delta_0$ are parameters fitted to measurements of energies of lower-lying states for each series from literature \cite{George1990c,Johansson1967,moore1949atomic}. This gave the behaviour of the $\delta_n$ values for increasing $n$.
The $a_{0,1,2,3}$ and $\delta_0$ values obtained for these series are given in Table~\ref{tab:IP}.
The value for the IP can be determined by using it as a free parameter to fit the $\delta_n$ values to this expression. 
The importance of measuring Rydberg states over a wide range of $n$, not just for high-lying $n$ states, for fitting the IP with the $\delta_n$ values is clearly seen in Figure~\ref{fig:simul_IP}~b), as the uncertainty in $\delta_n$ scales as ${n^*}^{-3}$.
Furthermore, deviations in the experimental $\delta_n$ from that expected by Expression \ref{rydberg:ritz} can be used to identify deficiencies in the energy level measurements being used to determine the IP, due to a susceptibility to stray electric fields (in principle avoided by separation of the field ionization from the laser excitation step in a collinear setup) or perturbing configurations lying above the IP.
Large perturbations were found in the 5s$^2$($n$)d~$^2$D series from the $^2$D term of the 5s5p$^2$ configuration \cite{Neijzen1981a}.
No statistically significant deviation outside of the values of Expression~\ref{rydberg:ritz} was observed within the accuracy of the 5s$^2$($n$)f~$^2$F$_{5/2}$, 5s$^2$($n$)f~$^2$F$_{5/2,7/2}$, 5s$^2$($n$)p~$^2$P$_{1/2}$, 5s$^2$($n$)p~$^2$P$_{3/2}$ series measurements performed in this work.

The $\delta_n$ values were fitted to Expression \ref{rydberg:ritz} for each Rydberg series to obtain the value for the IP independently for each.
The resulting IP values are presented in Table~\ref{tab:IP} and in the sub plots of Figure~\ref{fig:simul_IP}b).
The IP values extracted from the series measurements and from literature are in good agreement.
The $\delta_n$ obtained from the 5s$^2$($n$)p~$^2$P$_{1/2,3/2}$ and 5s$^2$($n$)s~$^2$S$_{1/2}$ series taken from literature \cite{Neijzen1982a, Neijzen1981a} are also shown in Figure~\ref{fig:simul_IP}b) and the resulting IP values in Table~\ref{tab:IP}. 

As the value of the IP is a common parameter for all of the series, a global simultaneous fit with the IP as a free parameter was performed using the $\delta_n$ values from the series measurements in this work in addition to literature, taking into the account the error of the individual $\delta_n$ values and reducing possible sources of systematic error. 
This yielded a combined value for the IP of \SI{46670.1055(21)}{\per\centi\meter}, an improvement over the previous highest precision literature value for the IP of \SI{46670.106(6)}{\per\centi\meter} (Ref.~\cite{Neijzen1982a} from the 5s$^2$($n$)p~$^2$P$_{1/2,3/2}$ series.

The difference of 0.2\% of the theoretical IP from the experimental value is well within that expected under the RCSSD approximation \cite{Das2011}, in contrast to difference by a factor of 5 observed in the $A_{\text{hf}}$ constants. This further highlights the difference in electron correlation trends for the calculation of hyperfine structure constants, in contrast to energies, for the same RCSSD level of approximation.

\begin{table}[h]
\centering
\begin{tabular}{|c|c|c|c|c|c|c|}
\hline
 Series & IP (cm$^{-1}$) & $\delta_0$ & $a_0$ & $a_1$ & $a_2$ & $a_3$ \\
\hline
\textbf{Theory} & &  &  &  &  & \\
DHF & 41507.11 &  &  &  &  & \\
RCCSD & 46762.85 &  &  &  &  & \\
$+$Breit & 46725.95 & \\
$+$QED   & \textbf{46763.57} & \\
\cmidrule{0-1}
Literature & &  &  &  &  & \\
($n$)p $^2$P$_{1/2, 3/2}$, Ref.\cite{Neijzen1982a} & 46670.106(6) &  &  &  &  & \\
($n$)s $^2$S$_{1/2}$, Ref.\cite{Neijzen1981a} & 46670.107(10) &  &  &  &  & \\
($n$)f $^2$F$_{5/2, 7/2}$, Ref.\cite{Johansson1967}, 7$\leq n \leq$10& 46670.110(50) &  &  &  &  & \\
\cmidrule{0-1}
This work & &  &  &  &  & \\
($n$)f $^2$F$_{5/2}$ & 46670.110(4) & 0.04 & 0.041 & -0.153 & 0.714 & -2.296 \\
($n$)f $^2$F$_{5/2, 7/2}$ & 46670.102(6) & 0.04 & 0.041 & -0.153 & 0.714 & -2.296 \\
($n$)p $^2$P$_{3/2}$ & 46670.105(4) & 3.22 & 3.196 & 0.382 & 0.125 & 3.1454 \\
($n$)p $^2$P$_{1/2}$  & 46670.106(7) & 3.25 & 3.223 & 0.380 & 0.112 & 3.1258 \\
\cmidrule{0-1}
\textbf{Global fit} &  &  &  &  &  & \\
(This work \&  & \textbf{46670.1055(21)} &  &  &  &  & \\
Literature) &  &  &  &  &  & \\
\hline
\end{tabular}
\caption{\label{tab:IP}Values for the ionization potential of the indium atom evaluated by fitting the the quantum defects of the 5s$^2$($n$)f $^2$F$_{5/2}$ and 5s$^2$($n$)p $^2$P$_{3/2,1/2}$ Rydberg series, fitting to the series separately and simultaneously (`sim. fit.'). The determined $a_0,1,2,3$ and $\delta_0$ parameters for Expression~\ref{rydberg:ritz} are presented. }
\end{table}

\FloatBarrier


\section*{\hypertarget{FIdesign}{Systematics of the field-ionization setup}}

\begin{figure}[h]
\centering
\includegraphics[width=17cm]{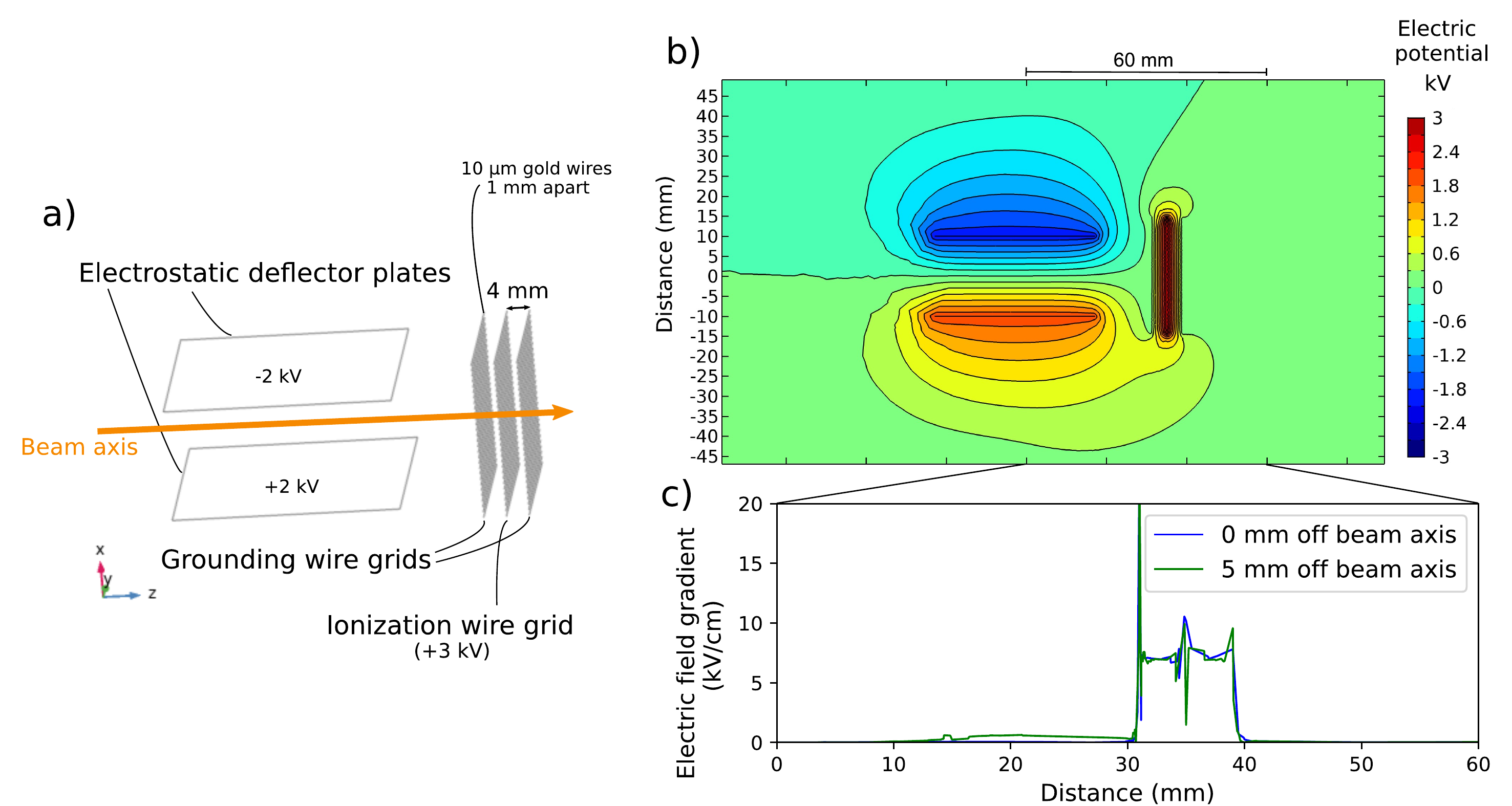}
\caption{a) A schematic of the electrode arrangement used in this work b) The electric potential of the field-ionization arrangement used for the measurements of this work, and c) the accompanying electric field gradient, simulated from measurements of voltages applied during this work.}
\label{fig:oldsim}
\end{figure}

The field ionization of the Rydberg states in this work was performed using the electrode arrangement shown in Figure~\ref{fig:oldsim}, located in the position indicated in Figure~\ref{fig:CRIS_layout_old}.
Three consecutive grids of parallel gold wires, of \SI{10}{\micro\meter} thickness and with \SI{1}{\milli\meter} separation between wires were used as electrodes to create the field for ionization, as shown in Figures~\ref{fig:oldsim}a), and \ref{fig:oldsim}b). The outmost grids were used to provide ground shielding.
The wire grids were mounted on a printed circuit board and spaced \SI{4}{\milli\meter} apart, resulting in an average electric field gradient of \SI{7.5}{\kilo\volt\per\centi\meter} for the 3~kV potential applied to the innermost grids.
The arrangement included two parallel electrostatic deflector plates with opposing polarity before the grids, in order to deflect background ions.
These background ions originate from non-resonant processes in the preceding \SI{120}{\centi\meter} flight path between the charge exchange cell and the field-ionization grids, which is referred to as the `laser-atom interaction region'.
The region surrounding the grids is further called the `field-ionization region', and the region between the last field-ionization grid and the ion detector will be referred to as the `post-ionization region'.
The regions are also indicated in Figure~\ref{fig:CRIS_layout_old}.

This longitudinal electric field-ionization arrangement was chosen so the closely spaced grids could provide a small ionization volume of \SI{0.4}{\centi\meter\cubed} with a very well defined electric field gradient.
This small volume has to be compared to the \SI{120}{\centi\meter\cubed} volume where non-resonant laser ionization would normally take place.
This is a reduction by a factor of 300 in volume in which collisional ionization of the atom beam can occur with residual gases.
This is a substantial source of background which can be removed when field ionization is used.
An additional advantage of this setup is that the grids approximate a plane geometry for the electric field, removing the dependence on transverse displacement of the atom beam on the field gradient experienced by the atoms.
This can be seen in Figure~\ref{fig:oldsim}c) where very difference in electric field gradient is observed on a \SI{1}{\milli\meter} level, outside of the grid regions, for an offset of \SI{5}{\milli\meter} from the beam axis (to simulate an incident atom beam of this width), compared to traditional tube field-ionization geometries where large differences in electric field gradient can be found transverse to the beam axis \cite{Aseyev1993b,Stratmann1994a}.
The electrostatic simulations were performed using COMSOL Multiphysics~$^{\circledR}$ \cite{AB}.

The spread in the position where the Rydberg atoms are ionized is determined by the ionization probability for the Rydberg atom in the electric field gradient created by the electric potential. Therefore the ionization probability in a given electric field gradient ultimately determines the spread in electric potential the ions are produced and the energy spread of the ion beam as it exits the `field-ionization region'.
The situation of the Rydberg atom bunch encountering a step increase in electric field as they travel into `field-ionization region' is equivalent to the application of a pulsed electric field to the Rydberg atoms at rest, which has been studied more extensively \cite{Robicheaux1997,Baranov1994,Jones1993}.
In the adiabatic limit where the classical electron motion is fast compared to the electric field pulse, the field necessary to reach saturation of ionization for the ensemble of Rydberg atoms (to ionize all Rydberg states above a given $n$ within the pulse duration) is calculated \cite{Baranov1994} to be $E_n - \text{IP} = -5.97F^{1/2}$~V/cm, corresponding to a field gradient of

\begin{equation}
    F_{sat}=3.38\times\frac{10^8}{n^{*4}} \; \text{V/cm},
    \label{eqn:adiabatic}
\end{equation}

using the parameters for an indium Rydberg atom. This is similar to the commonly used estimate for the critical field ionization strength in the case of a static electric field \cite{Stratmann1994a,Ducas1975} of $F_{crit} \approx 3\times10^8/n^{*4} \; \text{V/cm}$.
The classical Kepler period \cite{Jones1993} of
\begin{equation}
    \tau_K=2\pi\frac{m_e a_0^2}{\bar{h}} n^{*3}  \; \; ,
\end{equation}
for an electron in the 5s$^2$70f~$^2$F$_{5/2}$ state is $\tau_K$=\SI{52}{\pico\second}, where $a_0$ is the Bohr radius and $m_e$ the electron mass.
This can be used to estimate the cutoff for the adiabatic limit.
The distance within which Rydberg atoms can be assumed to be ionized applying the electric field gradients according to Expression \ref{eqn:adiabatic}, is then given by
$l_{sat}=\tau_K\nu_B$, for an atom beam of velocity $\nu_B$. In the case of atomic $^{115}$In at $T_\text{B}$~=~\SI{25}{\kilo\electronvolt} this corresponds to \SI{10.65}{\micro\meter}.
This results in a minimum energy spread of $\Delta E$~=~\SI{8}{\electronvolt} for the electric gradient of $F$=\SI{7.5}{\kilo\volt\per\centi\meter} used in this work ($\Delta E=l_{sat}F$).
The corresponding time-of-flight broadening for this minimum energy spread is well below the \SI{4}{\micro\second} ion bunch width from the ablation ion source and was not resolvable in this work.
In order to go below this intrinsic energy spread, higher electric field gradients would be required to ensure ionization in a short distance, although scaling as $F \propto n^{*-2}$ appears in the sub-ps regime \cite{Jones1993}.

\begin{figure}[h]
\centering
\includegraphics[width=18cm]{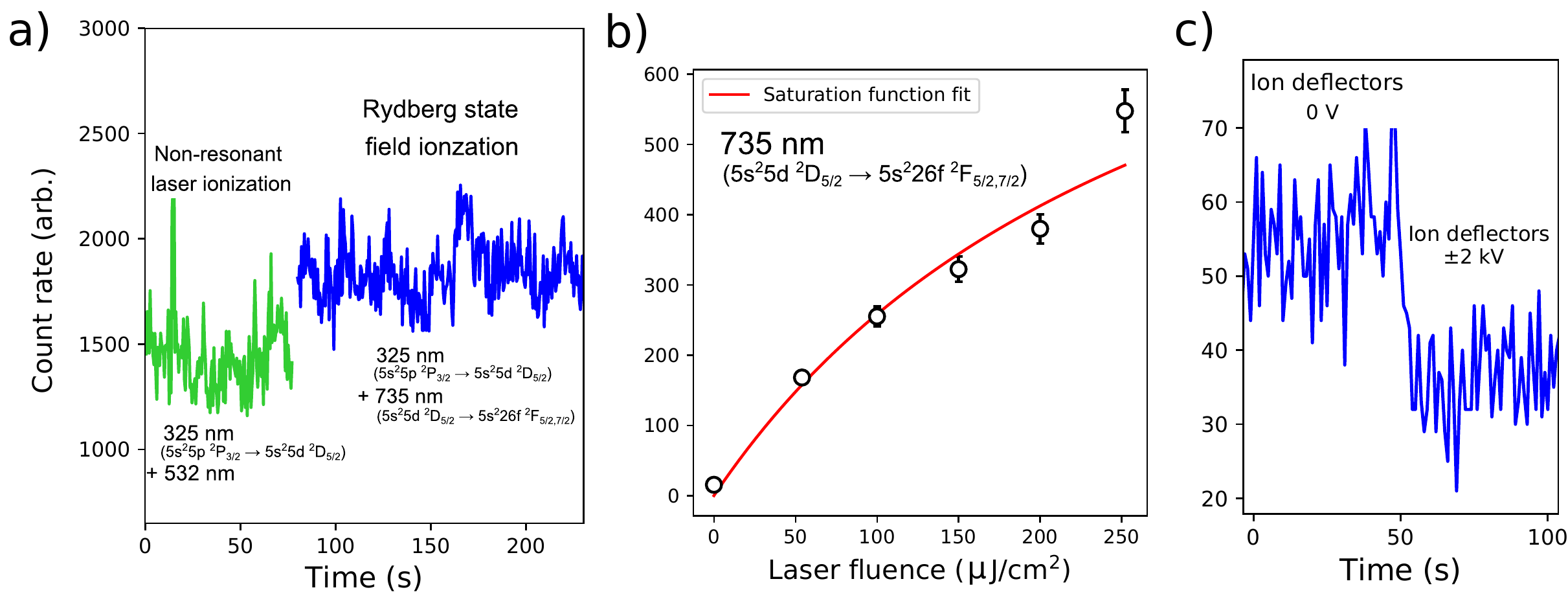}
\caption{a) The count rate on resonance for a typical non-resonant photoionization last step versus the field-ionization scheme corresponding to their total ionization efficiencies, b) fitting the saturation curve \cite{Siegman1986} of the 5s$^2$5d~$^2$D$_{3/2}$ $\rightarrow$ 5s$^2$26f~$^2$F$_{5/2,7/2}$ transition, giving a saturation fluence of 293(140)~$\mu$J/cm$^2$ and c) background rates with and without electrostatic deflectors to estimate background in the `interaction' and `post-ionization' region.}
\label{fig:532compar}
\end{figure}

Figure~\ref{fig:532compar}a) shows the detected ion-count rate of a measurement performed using the 5s$^2$5p~$^2$P$_{3/2}$ $\rightarrow$ 5s$^2$5d~$^2$D$_{5/2}$ (325.6~nm) first step followed by non-resonant ionization using 532-nm light produced by a Litron TRLi HR 120-200 Nd:YAG laser.
This was recorded in order to make a comparison with the total ionization efficiency of the field-ionization setup, using the same first step followed by the 5s$^2$5d~$^2$D$_{3/2}$ $\rightarrow$ 5s$^2$26f~$^2$F$_{5/2,7/2}$ transition and field ionization.
The measurements were performed less than a minute apart following optimization of the overlap of the 735-nm, 532-nm and 325-nm light with the neutral atoms, aligned using two irises before and after the interaction region of the beamline (as indicated in Figure~\ref{fig:CRIS_layout_old}).
The laser pulses were overlapped in time using a photodiode at the laser exit window of the beamline.
A maximum output pulse fluence of 55~mJ/cm$^{2}$ was used for the 532-nm step, with no discernible decrease in count rate observed down to 44~mJ/cm$^{2}$. 
Meanwhile the 5s$^2$5d~$^2$D$_{3/2}$~$\rightarrow$~5s$^2$26f~$^2$F$_{5/2,7/2}$ transition appeared to not be saturated as indicated in Figure \ref{fig:532compar}b), with a maximum of 250(20)~$\mu$J/cm$^2$ available and the estimated saturation fluence of 293(140)~$\mu$J/cm$^2$.
A scatter of around 10\% in beam intensity was due to shot-to-shot variations intrinsic to the ablation ion source setup used \cite{Wang2014, GarciaRuiz2018}.
While this makes an exhaustive study difficult, this underlines two issues under these typical measurement conditions:
i) The efficiency of a non-resonant laser ionization scheme is typically lower than a field-ionization setup \cite{Burnett1993}.
This can be due to the non-linear nature of non-resonant photoabsorption into the continuum \cite{Burnett1993}, reduced efficiency of collecting and detecting ions from a larger volume or re-neutralisation of the ions (at pressures of around \SI{1E-9}{\milli\bar} this can still give an appreciable contribution \cite{Vernon2019}).
ii) A larger laser fluence is required in order to saturate transitions to high $n$ Rydberg states such as the 5s$^2$5d~$^2$D$_{3/2}$~$\rightarrow$~5s$^2$26f~$^2$F$_{5/2,7/2}$ transition, as the transition strength decreases with $n$ \cite{Shahzada2012}.
An appropriate $n$ Rydberg state has to be chosen to ensure saturation of the transition in addition to a electric field gradient to ensure ionization according to Expression~\ref{eqn:adiabatic}.
This is an additional validation of the fact that the technique lends itself well to use on bunched atomic beams, where high laser fluence pulsed lasers can be used.\\

In order to study the factor of reduction in collisional background using the field-ionization setup shown in Figure~\ref{fig:oldsim}, measurements were performed at pressures raised by a factor of 10 compared to the nominal operating level, increasing the signal for the collisional background rate.
The pressure in the `interaction' region of length $l_1=$\SI{120}{\centi\meter} was raised to \SI{5E-9}{\milli\bar} ($\rho_1=$\SI{1.2E14}{\per\meter\cubed}), and the `post-ionization' region of length $l_2=$\SI{30}{\centi\meter} was raised to \SI{5E-8}{\milli\bar} ($\rho_2=$\SI{1.2E15}{\per\meter\cubed}).

For a measured neutral beam current of $I_B=$\SI{6.0(1)E6} atoms/s, and a collisional ion beam current, I$_C$, the cross section for collisional ionization can be defined as 
\begin{equation}
\sigma = \frac{I_{C}}{I_B \rho l} \; \; \text{cm}^2 \; .
\end{equation}

As both regions will have the same value of $\sigma$, measurements of the collisional ion currents can be used as a consistency check for the reduction in ionization volume using the known atom path lengths, $l$, and residual gas densities, $\rho$.
The remaining background ion current from applying $\pm$2~kV electrostatic deflectors in the `field-ionization' region gave the collisional ion current for the `post-ionization' region, while applying the ground potential gave the background ion current from both `interaction' plus `post-ionization' regions.
The measured ion currents were I$_{C}^{l2}$~=~35(5) ions/s and I$_{C}^{l1+l2}$~=~55(5) ions/s, respectively, as shown in Figure~\ref{fig:532compar}c).
From these measurements the collisional ionization cross sections for the indium atom incident upon residual gas atoms at 25~keV were determined to be $\sigma_1$~=~\SI{2.3(8)E-16}{\centi\meter\squared} and $\sigma_2=$\SI{1.62(23)E-16}{\centi\meter\squared} for the `interaction' and `post-ionization' regions, respectively.
The larger error of $\sigma_1$ results from taking the difference between I$_{C}^{l2}$ and I$_{C}^{l1+l2}$ to determine I$_{C}^{l1}$.
The cross sections are in agreement and are of the expected order of magnitude at a beam energy of 25~keV \cite{Kunc1991}.
This demonstrates a consistency for a factor of five in length (and volume assuming a homogeneous beam diameter) reduction (from 150~cm to 30~cm, as indicated in Figure~\ref{fig:CRIS_layout_old}) for the source of collisional background ions.
In addition this shows that the largest source of remaining atom-beam related background is due to ions created by collisional ionization with residual gases in the `post-ionization' region, which are not able to be removed by the electrostatic deflectors in the `field-ionization' region.
The background suppression of the design in this work is therefore limited by the length of the `post-ionization region' and the vacuum pressure in that region. 

The simulated electric field gradient of Figure \ref{fig:oldsim}c) highlights an additional consideration when using parallel wires for field ionization. The approximation of a planar electric potential breaks down as the wires are approached and inhomogenities in the penetrating field create a large spike in the experienced electric field gradient.
This property is in fact useful for defining the point of ionization and reducing the ion energy spread, however the potential geometry of Figure \ref{fig:oldsim}b) creates three positions where the electric field gradient is greatest and approximately equal in magnitude. It is therefore crucial for the critical field for ionization saturation to be applied to avoid ionization across more than one position which would result in a maximum energy spread of the magnitude of the potential applied.

\subsection*{An improved field-ionization setup}

Although the design used in this work effectively removes background from collisional ions created in the `interaction' region before the `field-ionization' region, the remaining background from ions created in the `post-ionization region' can still be substantial.

\begin{figure}[h]
\centering
\includegraphics[width=17cm]{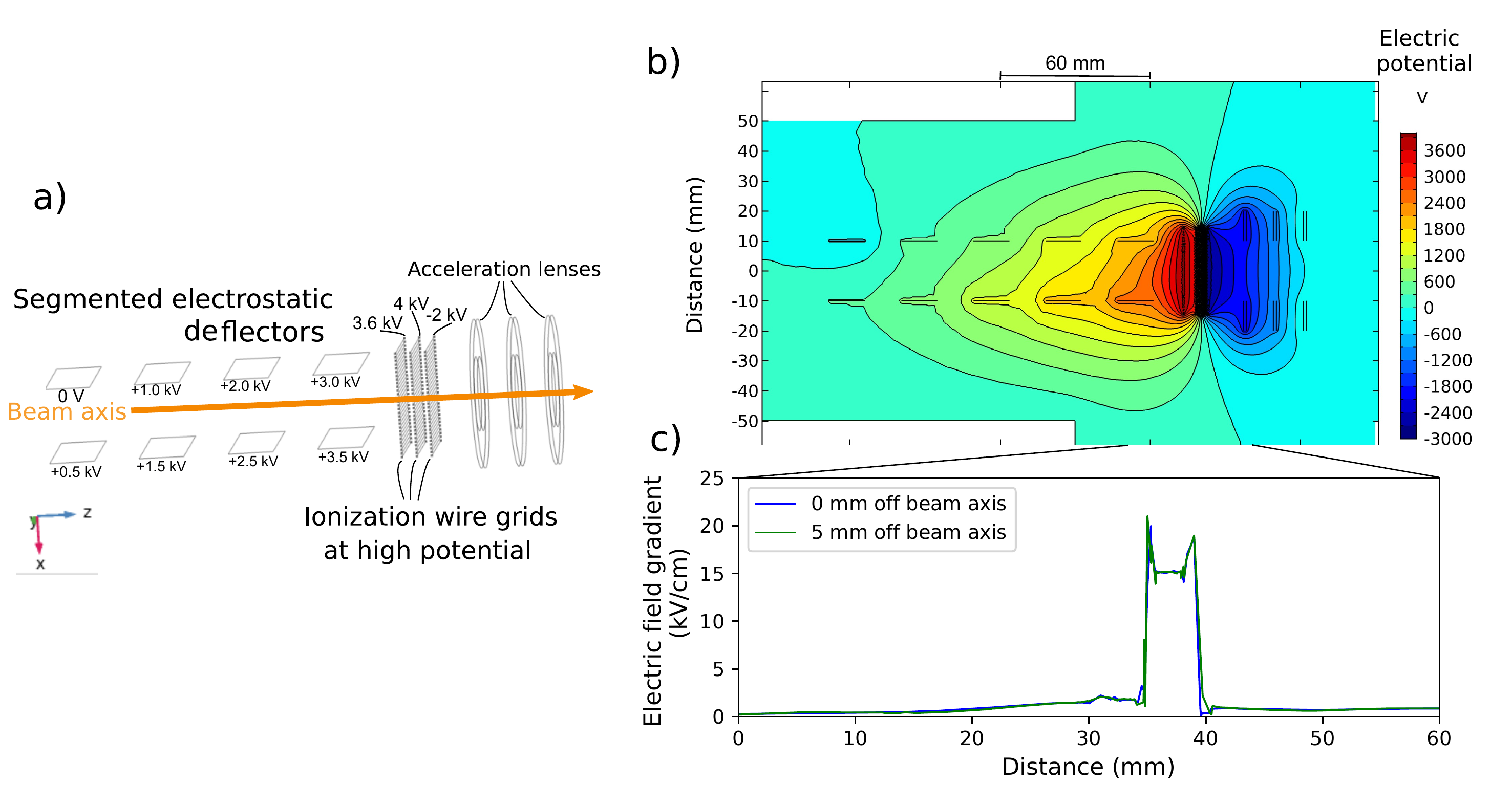}
\caption{a) A schematic of the improved electrode arrangement simulated from this work b) The electric potential of an optimised field-ionization setup, and b) the accompanying electric field gradient, simulated using voltages compatible with a 25~keV beam.}
\label{fig:newsim}
\end{figure}

With this consideration, an improved design has been developed to detect only those ions created inside the small volume of the field-ionization grids and is presented in Figure~\ref{fig:newsim}.
The principle of the design is to create an energy shift for the ions created in the `field-ionization' region.
This introduces energy selectivity for the Rydberg states ionized in the `field-ionization' region, distinguishing them from other background sources of ions which will remain at the initial beam energy.
Compared to the design used for the measurements in this work, this removes the demand for a short `post-ionization region' with the best possible vacuum conditions.

\begin{figure}[h]
\centering
\includegraphics[width=16cm]{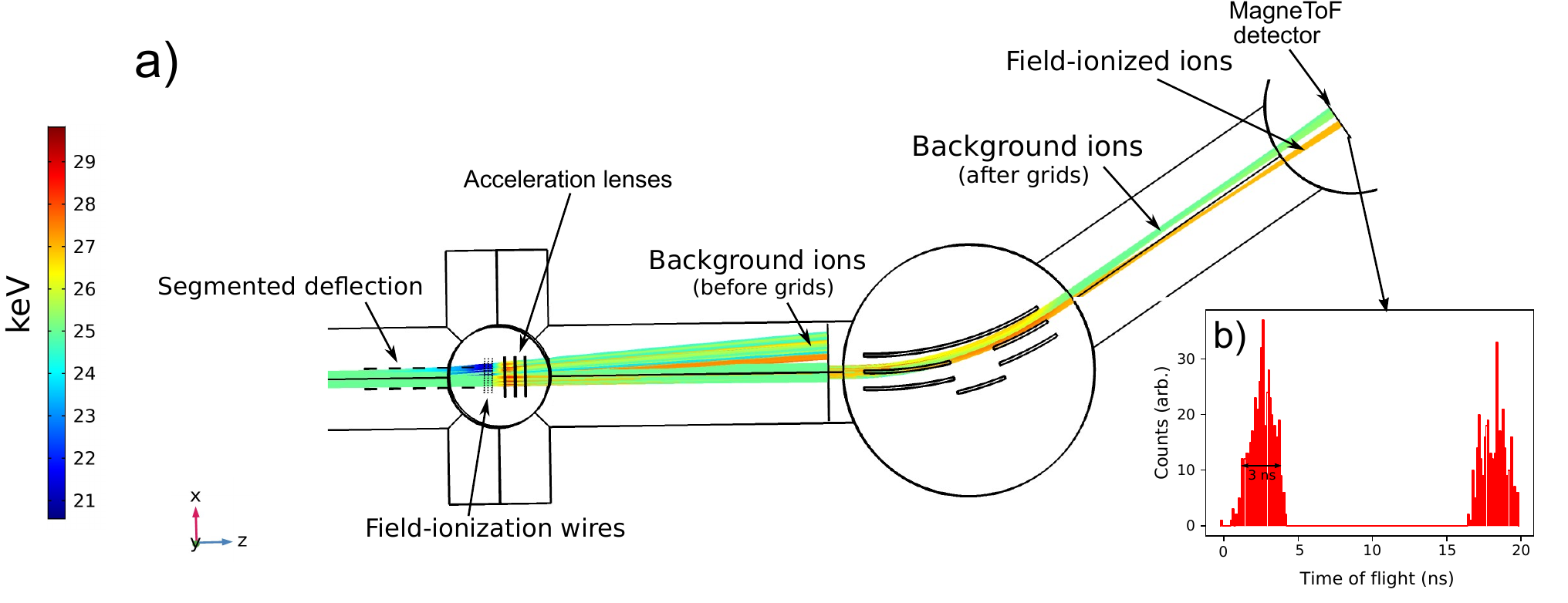}
\caption{a) The simulated ion trajectories using the field-ionization arrangement shown in Figure \ref{fig:newsim} and b) the corresponding simulated time-of-flight difference for ions from field ionization versus background.}
\label{fig:newpaths}
\end{figure}

In this improved design, the opposite polarity deflector plates (Figure \ref{fig:oldsim}a,b) ) are replaced by segmented flat electrodes \cite{Stratmann1994a} of the same polarity, but with a potential difference of around 500~V between them to provide the equivalent electrostatic deflection of background ions created before the field-ionization grids (Figure~\ref{fig:newpaths}a,b) ). 
These electrodes are labelled as ``segmeneted electrostatic deflectors'' in Figure~\ref{fig:newsim} a).
Electrostatic lenses are also included following the field-ionization grids, labelled as ``acceleration lenses'' in Figure~\ref{fig:newsim} a).
These additional ion optic elements were designed with simple planar geometries to be compatible with fabrication using metal traces on printed circuit boards \cite{Cooper2019}.

This design allows the outer grids to be held at a higher and adjustable electric potential without compromising the advantages of the previous field-ionization arrangement.
The segmented electrodes allow the Rydberg atoms to enter a high potential without an abrupt increase in electric field gradient causing ionization.
The removal of the opposite polarity deflector plates allows the potential of the first grid to be raised without introducing a large asymmetry in the electric potential or a large electric field gradient transverse to the atom beam axis.
In addition, this creates a well defined electric field gradient without the need for outer grounding grids.
The principle of this arrangement is to reduce the electric field gradient between the first and second grids, moving the step to high electric field gradient to the middle grid instead.
The step can be made greater by applying an opposite polarity potential to the last grid, as shown in Figure~\ref{fig:newsim}.
This localizes the field-ionization region to a raised potential, resulting in an increase in beam energy of Rydberg states which are ionized in this potential.
The `acceleration lenses' following the field-ionization grids, are then used for extraction from the raised potential.

The resulting ion trajectories from this ionization arrangement are shown in Figure~\ref{fig:newpaths}a), where the increase in beam energy of ions created inside the field-ionization region is indicated.
The ions with the beam energy of interest can be selectively detected following electrostatic deflection, because the deflection introduces an angular separation of ions with different energies, as seen in Figure~\ref{fig:newpaths}a).
Using slits (or a position sensitive detector) to select ions of a given beam energy, allows the ions created by field ionization to be distinguished from any other background source of ions created in the `interaction' or `post-ionization' regions.
This not only includes collisional ions, but ions created by field ionization of collisionally excited or re-neutralised atoms in the field of the \SI{20}{\degree} bend to the ion detector, photoionization or molecular breakup, as all of these sources of background ions will remain at the lower beam energy.
Alternatively the beam energy could be measured directly \cite{Harasimowicz2012}, or the difference in detected time of flight of the ions could be used as a gate if the bunch width was sufficiently narrow.
For example, the time of flight separation introduced in the `post-ionization' region for the ions travelling at 25~keV is around 15~ns (Figure~\ref{fig:newpaths}b), so a bunch window narrower than this would be needed.
The incident temporal atom bunch width of \SI{4}{\micro\second} in this work would prevent this.
The beam energy difference for ions created by field ionization could be enhanced by using lower incident beam energies or higher potential for the ionization apparatus, however the design of the ion optics then becomes more critical to avoid ion transmission loses.

The improved field-ionization design outlined here therefore offers improved background suppression over the design used for measurements in this work, by providing selectivity of ions created by field ionization independent of the length and vacuum quality of the `post-ionization' region.
In general, the background suppression factor for the improved field-ionization design compared to non-resonant laser ionization can be expressed as

\begin{equation}
    S_{BG} = \frac{L}{l_{ion}} \frac{P_L}{P_l} \; \; .
\label{BG_supp}
\end{equation}

where $L$ is the path length of the `interaction' plus 'post-ionization' regions, $l_{ion}$ is the path length in which ionization can take place inside the `field-ionization' region, and  $P_L/P_l$ is the ratio vacuum pressure in the two regions.
This is under the approximation of a homogeneous gas composition in the regions and a uniform atomic beam diameter.
The energy selectivity offers the prospect of a reduction in ionization volume by a factor of \SI{1.6E5}, down from a region of length $L$~=~\SI{150}{\centi\meter} to the $l_{ion}$~=~\SI{10.65}{\micro\meter} for the adiabatic cut-off assumed in the field-ionization model of Expression~\ref{eqn:adiabatic}, where $l_{ion}$~=~$l_{sat}$~=~$\tau_K\nu_B$.
However the electrostatic bend used in the CRIS experimental setup, combined with adjustable slits to select an ion path incident on the detector has an energy resolution of around $\sigma_ E$~=~\SI{1.5}{\kilo\electronvolt}, which can only guarantee a selectivity of the ionization volume down to $l_{ion}$~=~$F$/$\sigma_E$.
For the value of $F$~=~\SI{7.5}{\kilo\volt\per\centi\meter} used in this work, this corresponds to a volume reduction by a factor of \SI{1.25E3}{}.
Below this limit, direct beam energy measurement, or ion time-of-flight measurement using ion bunches narrower than 15~ns would be necessary to determine the actual energy spread and confirm the precise background suppression factor.

When combined with extreme-high vacuum technologies \cite{Thompson1976} to improve the vacuum quality in the field-ionization region (increase the $P_L/P_l$ ratio), this technique has the potential to reduce the dominating collisional background ion contribution to a vanishingly low level when compared to other sources of background, such as non-resonant ionization from the lower pulse energy resonant step laser light, the dark count rate of the detector ($\sim$0.08 cps for an ETP DM291 MagneTOF), or residual radioactivity in the setup.\\

\section*{Conclusion}

The use of ion cooling and bunching has allowed the most sensitive measurements of exotic atoms and molecules containing short-lived isotopes to date \cite{DeGroote2020,Ruiz2019}, by concentrating measurements on ion bunches  into a narrow time window, in order to improve background suppression and additionally allowing a high duty cycle for high-resolution and high-detection efficiency pulsed laser ionization spectroscopy \cite{Koszorus2019,Vernon2019a,Flanagan2013}.

In this work we have implemented field ionization with the Collinear Resonance Ionization Spectroscopy (CRIS) technique, to further increase the selectivity (and thus sensitivity) of high-resolution measurements of hyperfine spectra of isotopes in atom bunches.
This allows the ionization to take place in a narrow spatial window in addition to the narrow time window, substantially reducing background due to collisional ions created alongside the atoms of interest in larger ionization volumes. Here we have demonstrated a factor of five in ionization volume reduction and corresponding background suppression, when accounting for vacuum pressure.
In principle this will allow measurements of exotic isotopes with yields down to 4 atoms per second at the CRIS experiment.
However, a further factor of $>$400 improvement in background suppression of collisional ionization shown to be possible with an improved design, which also makes background suppression independent of distance from field ionization to ion detection by incorporating an increase in beam energy of the field-ionized Rydberg atoms.
Furthermore, as a non-resonant pulsed laser step is no longer necessary to ionize the atom bunches, this removes a significant source of photo-ionization background, in addition to removing a source of AC Stark shifts in measurements from short-lived metastable states \cite{Koszorus2019a}.

By using bunched atomic beams the technique is well suited to the use of narrow-band pulsed lasers, taking advantage of the high spectral density to saturate transitions to high-lying Rydberg states required for field ionization.
The 5s$^2$($n$)p~$^2$P and 5s$^2$($n$)f~$^2$F Rydberg series states in the indium atom up to $n$~=~72 were studied and used to evaluate the ionization potential of the indium atom to be \SI{46670.1055(21)}{\per\centi\meter}, in agreement with, and improving upon the precision of previous measurements.
Furthermore, the technique allows high resolution measurements of the hyperfine structure constants and isotope shifts of individual atomic states directly.

The nuclear magnetic dipole, nuclear electric quadrupole hyperfine structure parameters and isotope shifts of the $^{113}$In and $^{115}$In isotopes, for the 5s$^2$5d $^2$D$_{5/2}$ and 5s$^2$5d $^2$D$_{3/2}$ states were measured.
The experimental results were compared to DHF, RCCSD and AR-RCCSD calculations, where a good level of agreement was found with experimental isotope shifts and the ionization potential of the indium atom.
While the RCCSD calculations showed an improvement over DHF calculations for the A$_\text{hf}$ constants, giving the correct signs, the magnitudes were underestimated, indicating that electron correlations play a crucial role in the hyperfine stucture constants of these 5s$^2$5d~$^2$D$_{5/2}$ and 5s$^2$5d~$^2$D$_{3/2}$ states, demanding further theoretical study.

Improvements in highly sensitive detection techniques compatible with precise laser spectroscopy are required to measure the nuclear structure of the most exotic nuclei produced at radioactive beam facilities, important for developing nuclear theories \cite{Stroberg2019a,Ekstrom2019a,Morris2018}, in addition to paving the way for measurement of new observables from atomic nuclei when combined with precise calculations of the parameters of atomic states and their symmetries \cite{Reinhard2019a, Flambaum2019a, Berengut2018b, Carlson2015b}.
Furthermore, high sensitivity ionization of isotopes has many potential applications, such as the separation of nuclear waste \cite{Fujiwara2019}, enrichment of nuclear fuel \cite{Paisner1988}, collection of nuclear isomers \cite{Walker1999}, ``ultra''-trace analysis \cite{Lu2003}, research of nuclear-spin-dependent effects \cite{Swift2018,Kane1998a} and highly-purified nuclear decay spectroscopy \cite{Lynch2014}.

\section*{Acknowledgements}

This work was supported by ERC Consolidator Grant No.648381 (FNPMLS); 
STFC grants ST/L005794/1, ST/L005786/1, ST/P004423/1 and Ernest Rutherford Grant No. ST/L002868/1; 
GOA 15/010 from KU Leuven; the FWO-Vlaanderen (Belgium); 
National Key R\&D Program of China (Contract No: 2018YFA0404403),
the National Natural Science Foundation of China (No:11875073).
We would also like to thank A.J.~Smith and The University of Manchester workshop for their support. B.K.S. acknowledges use of the Vikram-100 HPC cluster of Physical Research Laboratory, Ahmedabad for carrying out atomic calculations. 

\section*{Author contributions statement}

A.R.V and K.T.F. conceived the experiment(s).
A.R.V, C.M.R, B.S.C., F.P.G, Q.W, R.F.G., H.A.P., F.J.W conducted the experiment(s).
A.R.V and C.M.R analysed the results.
A.R.V and B.K.S. prepared the manuscript.
B.K.S performed the coupled cluster method atomic calculations.
J.B., T.E.C., G.N., R.F.G., K.T.F and X.Y. provided advice, supervision and/or equipment.
All authors reviewed the manuscript.

\section*{Competing interests}

The authors declare no competing interests.

\section*{Additional information}

...

\FloatBarrier

\bibliography{main}

\begin{thebibliography}{10}
\urlstyle{rm}
\expandafter\ifx\csname url\endcsname\relax
  \def\url#1{\texttt{#1}}\fi
\expandafter\ifx\csname urlprefix\endcsname\relax\def\urlprefix{URL }\fi
\expandafter\ifx\csname doiprefix\endcsname\relax\def\doiprefix{DOI: }\fi
\providecommand{\bibinfo}[2]{#2}
\providecommand{\eprint}[2][]{\url{#2}}

\bibitem{Ruiz2019}
\bibinfo{author}{{Garcia Ruiz}, R.~F.} \emph{et~al.}
\newblock \bibinfo{journal}{\bibinfo{title}{{Spectroscopy of short-lived
  radioactive molecules: A sensitive laboratory for new physics}}}.
\newblock {\emph{\JournalTitle{arXiv preprint}}}  (\bibinfo{year}{2019}).
\newblock \eprint{1910.13416}.

\bibitem{Baumann2007a}
\bibinfo{author}{Baumann, T.} \emph{et~al.}
\newblock \bibinfo{journal}{\bibinfo{title}{{Discovery of
  {\$}{\^{}}{\{}40{\}}{\$}Mg and {\$}{\^{}}{\{}42{\}}{\$}Al suggests neutron
  drip-line slant towards heavier isotopes}}}.
\newblock {\emph{\JournalTitle{Nature}}} \textbf{\bibinfo{volume}{449}},
  \bibinfo{pages}{1022--1024}, \doiprefix\url{10.1038/nature06213}
  (\bibinfo{year}{2007}).

\bibitem{DeGroote2020}
\bibinfo{author}{de~Groote, R.~P.} \emph{et~al.}
\newblock \bibinfo{journal}{\bibinfo{title}{{Measurement and microscopic
  description of odd–even staggering of charge radii of exotic copper
  isotopes}}}.
\newblock {\emph{\JournalTitle{Nature Physics}}} \bibinfo{pages}{1--5},
  \doiprefix\url{10.1038/s41567-020-0868-y} (\bibinfo{year}{2020}).

\bibitem{Wienholtz2013}
\bibinfo{author}{Wienholtz, F.} \emph{et~al.}
\newblock \bibinfo{journal}{\bibinfo{title}{{Masses of exotic calcium isotopes
  pin down nuclear forces}}}.
\newblock {\emph{\JournalTitle{Nature}}} \textbf{\bibinfo{volume}{498}},
  \bibinfo{pages}{346--349}, \doiprefix\url{10.1038/nature12226}
  (\bibinfo{year}{2013}).

\bibitem{Fujiwara2019}
\bibinfo{author}{Fujiwara, T.}, \bibinfo{author}{Kobayashi, T.} \&
  \bibinfo{author}{Midorikawa, K.}
\newblock \bibinfo{journal}{\bibinfo{title}{{Selective Resonance
  Photoionization of Odd Mass Zirconium Isotopes Towards Efficient Separation
  of Radioactive Waste}}}.
\newblock {\emph{\JournalTitle{Scientific Reports}}}
  \textbf{\bibinfo{volume}{9}}, \doiprefix\url{10.1038/s41598-018-38423-4}
  (\bibinfo{year}{2019}).

\bibitem{Lynch2014}
\bibinfo{author}{Lynch, K.~M.} \emph{et~al.}
\newblock \bibinfo{journal}{\bibinfo{title}{{Decay-assisted laser spectroscopy
  of neutron-deficient francium}}}.
\newblock {\emph{\JournalTitle{Physical Review X}}}
  \textbf{\bibinfo{volume}{4}}, \bibinfo{pages}{1--15},
  \doiprefix\url{10.1103/PhysRevX.4.011055} (\bibinfo{year}{2014}).
\newblock \eprint{1402.4266}.

\bibitem{Lu2003}
\bibinfo{author}{Lu, Z.~T.} \& \bibinfo{author}{Wendt, K.~D.}
\newblock \bibinfo{title}{{Laser-based methods for ultrasensitive trace-isotope
  analyses}}, \doiprefix\url{10.1063/1.1535232} (\bibinfo{year}{2003}).

\bibitem{Swift2018}
\bibinfo{author}{Swift, M.~W.}, \bibinfo{author}{{Van De Walle}, C.~G.} \&
  \bibinfo{author}{Fisher, M.~P.}
\newblock \bibinfo{journal}{\bibinfo{title}{{Posner molecules: From atomic
  structure to nuclear spins}}}.
\newblock {\emph{\JournalTitle{Physical Chemistry Chemical Physics}}}
  \textbf{\bibinfo{volume}{20}}, \bibinfo{pages}{12373--12380},
  \doiprefix\url{10.1039/c7cp07720c} (\bibinfo{year}{2018}).
\newblock \eprint{1711.05899}.

\bibitem{Walker1999}
\bibinfo{author}{Walker, P.} \& \bibinfo{author}{Dracoulis, G.}
\newblock \bibinfo{journal}{\bibinfo{title}{{Energy traps in atomic nuclei}}}.
\newblock {\emph{\JournalTitle{Nature}}} \textbf{\bibinfo{volume}{399}},
  \bibinfo{pages}{35--40} (\bibinfo{year}{1999}).

\bibitem{Kane1998a}
\bibinfo{author}{Kane, B.~E.}
\newblock \bibinfo{journal}{\bibinfo{title}{{A silicon-based nuclear spin
  quantum computer}}}.
\newblock {\emph{\JournalTitle{Nature}}} \textbf{\bibinfo{volume}{393}},
  \bibinfo{pages}{133--137}, \doiprefix\url{10.1038/30156}
  (\bibinfo{year}{1998}).

\bibitem{Paisner1988}
\bibinfo{author}{Paisner, J.~A.}
\newblock \bibinfo{journal}{\bibinfo{title}{{Atomic vapor laser isotope
  separation}}}.
\newblock {\emph{\JournalTitle{Applied Physics B Photophysics and Laser
  Chemistry}}} \textbf{\bibinfo{volume}{46}}, \bibinfo{pages}{253--260},
  \doiprefix\url{10.1007/BF00692883} (\bibinfo{year}{1988}).

\bibitem{Miller2019b}
\bibinfo{author}{Miller, A.~J.} \emph{et~al.}
\newblock \bibinfo{journal}{\bibinfo{title}{{Proton superfluidity and charge
  radii in proton-rich calcium isotopes}}}.
\newblock {\emph{\JournalTitle{Nature Publishing Group}}}
  \textbf{\bibinfo{volume}{15}}, \bibinfo{pages}{432--436},
  \doiprefix\url{10.1038/s41567-019-0416-9} (\bibinfo{year}{2019}).

\bibitem{GarciaRuiz2016}
\bibinfo{author}{{Garcia Ruiz}, R.~F.} \emph{et~al.}
\newblock \bibinfo{journal}{\bibinfo{title}{{Unexpectedly large charge radii of
  neutron-rich calcium isotopes}}}.
\newblock {\emph{\JournalTitle{Nature Physics}}} \textbf{\bibinfo{volume}{12}},
  \bibinfo{pages}{594--598}, \doiprefix\url{10.1038/nphys3645}
  (\bibinfo{year}{2016}).
\newblock \eprint{arXiv:1602.07906v1}.

\bibitem{Wing1976}
\bibinfo{author}{Wing, W.~H.}, \bibinfo{author}{Ruff, G.~A.},
  \bibinfo{author}{Lamb, W.~E.} \& \bibinfo{author}{Spezeski, J.~J.}
\newblock \bibinfo{journal}{\bibinfo{title}{{Observation of the Infrared
  Spectrum of the Hydrogen Molecular Ion H D +}}}.
\newblock {\emph{\JournalTitle{Physical Review Letters}}}
  \textbf{\bibinfo{volume}{36}}, \bibinfo{pages}{1488--1491},
  \doiprefix\url{10.1103/PhysRevLett.36.1488} (\bibinfo{year}{1976}).

\bibitem{Campbell2016}
\bibinfo{author}{Campbell, P.}, \bibinfo{author}{Moore, I.~D.} \&
  \bibinfo{author}{Pearson, M.~R.}
\newblock \bibinfo{journal}{\bibinfo{title}{{Laser spectroscopy for nuclear
  structure physics}}}.
\newblock {\emph{\JournalTitle{Progress in Particle and Nuclear Physics}}}
  \textbf{\bibinfo{volume}{86}}, \bibinfo{pages}{127--180},
  \doiprefix\url{10.1016/j.ppnp.2015.09.003} (\bibinfo{year}{2016}).
\newblock \eprint{arXiv:1011.1669v3}.

\bibitem{Neugart2017}
\bibinfo{author}{Neugart, R.} \emph{et~al.}
\newblock \bibinfo{journal}{\bibinfo{title}{{Collinear laser spectroscopy at
  ISOLDE: new methods and highlights}}}.
\newblock {\emph{\JournalTitle{Journal of Physics G: Nuclear and Particle
  Physics}}} \textbf{\bibinfo{volume}{44}}, \bibinfo{pages}{064002},
  \doiprefix\url{10.1088/1361-6471/aa6642} (\bibinfo{year}{2017}).

\bibitem{Voss2016}
\bibinfo{author}{Voss, A.} \emph{et~al.}
\newblock \bibinfo{journal}{\bibinfo{title}{{The Collinear Fast Beam laser
  Spectroscopy (CFBS) experiment at Triumf}}}.
\newblock {\emph{\JournalTitle{Nuclear Instruments and Methods in Physics
  Research, Section A: Accelerators, Spectrometers, Detectors and Associated
  Equipment}}} \textbf{\bibinfo{volume}{811}}, \bibinfo{pages}{57--69},
  \doiprefix\url{10.1016/j.nima.2015.11.145} (\bibinfo{year}{2016}).

\bibitem{Minamisono2013}
\bibinfo{author}{Minamisono, K.} \emph{et~al.}
\newblock \bibinfo{journal}{\bibinfo{title}{{Commissioning of the collinear
  laser spectroscopy system in the BECOLA facility at NSCL}}}.
\newblock {\emph{\JournalTitle{Nuclear Instruments and Methods in Physics
  Research, Section A: Accelerators, Spectrometers, Detectors and Associated
  Equipment}}} \textbf{\bibinfo{volume}{709}}, \bibinfo{pages}{85--94},
  \doiprefix\url{10.1016/j.nima.2013.01.038} (\bibinfo{year}{2013}).

\bibitem{Flanagan2013}
\bibinfo{author}{Flanagan, K.~T.} \emph{et~al.}
\newblock \bibinfo{journal}{\bibinfo{title}{{Collinear Resonance Ionization
  Spectroscopy of Neutron-Deficient Francium Isotopes}}}.
\newblock {\emph{\JournalTitle{Physical Review Letters}}}
  \textbf{\bibinfo{volume}{111}}, \bibinfo{pages}{212501},
  \doiprefix\url{10.1103/PhysRevLett.111.212501} (\bibinfo{year}{2013}).

\bibitem{Kudriavtsev1982a}
\bibinfo{author}{Kudriavtsev, Y.~A.} \& \bibinfo{author}{Letokhov, V.~S.}
\newblock \bibinfo{journal}{\bibinfo{title}{{Laser method of highly selective
  detection of rare radioactive isotopes through multistep photoionization of
  accelerated atoms}}}.
\newblock {\emph{\JournalTitle{Applied Physics B Photophysics and Laser
  Chemistry}}} \textbf{\bibinfo{volume}{29}}, \bibinfo{pages}{219--221},
  \doiprefix\url{10.1007/BF00688671} (\bibinfo{year}{1982}).

\bibitem{Catherall2017}
\bibinfo{author}{Catherall, R.} \emph{et~al.}
\newblock \bibinfo{journal}{\bibinfo{title}{{The ISOLDE facility}}}.
\newblock {\emph{\JournalTitle{Journal of Physics G: Nuclear and Particle
  Physics}}} \textbf{\bibinfo{volume}{44}}, \bibinfo{pages}{094002},
  \doiprefix\url{10.1088/1361-6471/aa7eba} (\bibinfo{year}{2017}).

\bibitem{Ricketts2019}
\bibinfo{author}{Ricketts, C.~M.} \emph{et~al.}
\newblock \bibinfo{journal}{\bibinfo{title}{{A compact linear Paul trap cooler
  buncher for CRIS}}}.
\newblock {\emph{\JournalTitle{Nuclear Instruments and Methods in Physics
  Research Section B: Beam Interactions with Materials and Atoms}}}
  \doiprefix\url{10.1016/J.NIMB.2019.04.054} (\bibinfo{year}{2019}).

\bibitem{Mane2009}
\bibinfo{author}{Man{\'{e}}, E.} \emph{et~al.}
\newblock \bibinfo{journal}{\bibinfo{title}{{An ion cooler-buncher for
  high-sensitivity collinear laser spectroscopy at ISOLDE}}}.
\newblock {\emph{\JournalTitle{European Physical Journal A}}}
  \textbf{\bibinfo{volume}{42}}, \bibinfo{pages}{503--507},
  \doiprefix\url{10.1140/epja/i2009-10828-0} (\bibinfo{year}{2009}).

\bibitem{DeGroote2017b}
\bibinfo{author}{{De Groote}, R.} \emph{et~al.}
\newblock \bibinfo{journal}{\bibinfo{title}{{Dipole and quadrupole moments of
  Cu 73-78 as a test of the robustness of the Z=28 shell closure near Ni 78}}}.
\newblock {\emph{\JournalTitle{Physical Review C}}}
  \textbf{\bibinfo{volume}{96}}, \bibinfo{pages}{1--6},
  \doiprefix\url{10.1103/PhysRevC.96.041302} (\bibinfo{year}{2017}).

\bibitem{Dinger1986}
\bibinfo{author}{Dinger, U.} \emph{et~al.}
\newblock \bibinfo{journal}{\bibinfo{title}{{Collinear two-photon excitation of
  indium rydberg states in a fast atomic beam}}}.
\newblock {\emph{\JournalTitle{Zeitschrift fur Physik D Atoms, Molecules and
  Clusters}}} \textbf{\bibinfo{volume}{1}}, \bibinfo{pages}{137--138},
  \doiprefix\url{10.1007/BF01384668} (\bibinfo{year}{1986}).

\bibitem{Aseyev1993b}
\bibinfo{author}{Aseyev, S.~A.}, \bibinfo{author}{Kudryavtsev, Y.~A.} \&
  \bibinfo{author}{Petrunin, V.~V.}
\newblock \bibinfo{journal}{\bibinfo{title}{{Ionization of Fast Rydberg Atoms
  in Longitudinal and Transverse Electric Fields}}}.
\newblock {\emph{\JournalTitle{Appl. Phys. B}}} \textbf{\bibinfo{volume}{56}},
  \bibinfo{pages}{391--398} (\bibinfo{year}{1993}).

\bibitem{Vernon2019}
\bibinfo{author}{Vernon, A.} \emph{et~al.}
\newblock \bibinfo{journal}{\bibinfo{title}{{Simulation of the relative atomic
  populations of elements 1$\leq$Z$\leq$89 following charge exchange tested
  with collinear resonance ionization spectroscopy of indium}}}.
\newblock {\emph{\JournalTitle{Spectrochimica Acta Part B: Atomic
  Spectroscopy}}} \textbf{\bibinfo{volume}{153}}, \bibinfo{pages}{61--83},
  \doiprefix\url{10.1016/J.SAB.2019.02.001} (\bibinfo{year}{2019}).

\bibitem{GarciaRuiz2018}
\bibinfo{author}{{Garcia Ruiz}, R.~F.} \emph{et~al.}
\newblock \bibinfo{journal}{\bibinfo{title}{{High-Precision Multiphoton
  Ionization of Accelerated Laser-Ablated Species}}}.
\newblock {\emph{\JournalTitle{Physical Review X}}}
  \textbf{\bibinfo{volume}{8}}, \doiprefix\url{10.1103/PhysRevX.8.041005}
  (\bibinfo{year}{2018}).

\bibitem{Kopfermann1958}
\bibinfo{author}{Kopfermann, H.}
\newblock \emph{\bibinfo{title}{{Nuclear moments}}}
  (\bibinfo{publisher}{Academic Press}, \bibinfo{year}{1958}).

\bibitem{Niemax1980}
\bibinfo{author}{Niemax, K.} \& \bibinfo{author}{Pendrill, L.~R.}
\newblock \bibinfo{journal}{\bibinfo{title}{{Isotope shifts of individual nS
  and nD levels of atomic potassium}}}.
\newblock {\emph{\JournalTitle{Journal of Physics B: Atomic and Molecular
  Physics}}} \textbf{\bibinfo{volume}{13}}, \bibinfo{pages}{L461--L465},
  \doiprefix\url{10.1088/0022-3700/13/15/001} (\bibinfo{year}{1980}).

\bibitem{Sahoo2018}
\bibinfo{author}{Sahoo, B.} \& \bibinfo{author}{Das, B.}
\newblock \bibinfo{journal}{\bibinfo{title}{{Relativistic Normal
  Coupled-Cluster Theory for Accurate Determination of Electric Dipole Moments
  of Atoms: First Application to the Hg 199 Atom}}}.
\newblock {\emph{\JournalTitle{Physical Review Letters}}}
  \textbf{\bibinfo{volume}{120}}, \bibinfo{pages}{203001},
  \doiprefix\url{10.1103/PhysRevLett.120.203001} (\bibinfo{year}{2018}).

\bibitem{Sahoo2020}
\bibinfo{author}{Sahoo, B.~K.} \emph{et~al.}
\newblock \bibinfo{journal}{\bibinfo{title}{{Analytic response relativistic
  coupled-cluster theory: the first application to indium isotope shifts}}}.
\newblock {\emph{\JournalTitle{New Journal of Physics}}}
  \textbf{\bibinfo{volume}{22}}, \bibinfo{pages}{012001},
  \doiprefix\url{10.1088/1367-2630/ab66dd} (\bibinfo{year}{2020}).

\bibitem{Berengut2018b}
\bibinfo{author}{Berengut, J.~C.} \emph{et~al.}
\newblock \bibinfo{journal}{\bibinfo{title}{{Probing New Long-Range
  Interactions by Isotope Shift Spectroscopy}}}.
\newblock {\emph{\JournalTitle{Physical Review Letters}}}
  \textbf{\bibinfo{volume}{120}}, \bibinfo{pages}{091801},
  \doiprefix\url{10.1103/PhysRevLett.120.091801} (\bibinfo{year}{2018}).

\bibitem{Flambaum2018a}
\bibinfo{author}{Flambaum, V.~V.}, \bibinfo{author}{Geddes, A.~J.} \&
  \bibinfo{author}{Viatkina, A.~V.}
\newblock \bibinfo{journal}{\bibinfo{title}{{Isotope shift, nonlinearity of
  King plots, and the search for new particles}}}.
\newblock {\emph{\JournalTitle{Physical Review A}}}
  \textbf{\bibinfo{volume}{97}}, \bibinfo{pages}{032510},
  \doiprefix\url{10.1103/PhysRevA.97.032510} (\bibinfo{year}{2018}).

\bibitem{Reinhard2019a}
\bibinfo{author}{Reinhard, P.~G.}, \bibinfo{author}{Nazarewicz, W.} \&
  \bibinfo{author}{{Garcia Ruiz}, R.~F.}
\newblock \bibinfo{journal}{\bibinfo{title}{{Beyond the charge radius: the
  information content of the fourth radial moment}}}.
\newblock {\emph{\JournalTitle{arXiv preprint}}}  (\bibinfo{year}{2019}).
\newblock \eprint{1911.00699}.

\bibitem{Allehabi2020}
\bibinfo{author}{Allehabi, S.~O.}, \bibinfo{author}{Dzuba, V.~A.},
  \bibinfo{author}{Flambaum, V.~V.}, \bibinfo{author}{Afanasjev, A.~V.} \&
  \bibinfo{author}{Agbemava, S.~E.}
\newblock \bibinfo{journal}{\bibinfo{title}{{Using isotope shift for testing
  nuclear theory: the case of nobelium isotopes}}}.
\newblock {\emph{\JournalTitle{arXiv preprint}}}  (\bibinfo{year}{2020}).
\newblock \eprint{2001.09422}.

\bibitem{Flambaum2019a}
\bibinfo{author}{Flambaum, V.~V.} \& \bibinfo{author}{Dzuba, V.~A.}
\newblock \bibinfo{journal}{\bibinfo{title}{{Sensitivity of the isotope shift
  to the distribution of nuclear charge density}}}.
\newblock {\emph{\JournalTitle{Physical Review A}}}
  \textbf{\bibinfo{volume}{100}}, \doiprefix\url{10.1103/PhysRevA.100.032511}
  (\bibinfo{year}{2019}).
\newblock \eprint{1907.07435}.

\bibitem{Vernon2019a}
\bibinfo{author}{Vernon, A.~R.} \emph{et~al.}
\newblock \bibinfo{journal}{\bibinfo{title}{{Optimising the Collinear Resonance
  Ionisation Spectroscopy (CRIS) experiment at CERN-ISOLDE}}}.
\newblock {\emph{\JournalTitle{Nuclear Instruments and Methods in Physics
  Research Section B: Beam Interactions with Materials and Atoms}}}
  \doiprefix\url{10.1016/J.NIMB.2019.04.049} (\bibinfo{year}{2019}).

\bibitem{Koszorus2019}
\bibinfo{author}{Koszor{\'{u}}s, A.} \emph{et~al.}
\newblock \bibinfo{journal}{\bibinfo{title}{{Resonance ionization schemes for
  high resolution and high efficiency studies of exotic nuclei at the CRIS
  experiment}}}.
\newblock {\emph{\JournalTitle{Nuclear Instruments and Methods in Physics
  Research, Section B: Beam Interactions with Materials and Atoms}}}
  \doiprefix\url{10.1016/j.nimb.2019.04.043} (\bibinfo{year}{2019}).

\bibitem{Civis2014}
\bibinfo{author}{Civi{\v{s}}, S.} \emph{et~al.}
\newblock \bibinfo{journal}{\bibinfo{title}{{Laser ablation of an indium
  target: Time-resolved Fourier-transform infrared spectra of in i in the
  700-7700 cm-1range}}}.
\newblock {\emph{\JournalTitle{Journal of Analytical Atomic Spectrometry}}}
  \textbf{\bibinfo{volume}{29}}, \bibinfo{pages}{2275--2283},
  \doiprefix\url{10.1039/c4ja00123k} (\bibinfo{year}{2014}).

\bibitem{Wendt2000}
\bibinfo{author}{Wendt, K.}, \bibinfo{author}{Trautmann, N.} \&
  \bibinfo{author}{Bushaw, B.~A.}
\newblock \bibinfo{journal}{\bibinfo{title}{{Resonant laser ionization mass
  spectrometry: An alternative to AMS?}}}
\newblock {\emph{\JournalTitle{Nuclear Instruments and Methods in Physics
  Research, Section B: Beam Interactions with Materials and Atoms}}}
  \textbf{\bibinfo{volume}{172}}, \bibinfo{pages}{162--169},
  \doiprefix\url{10.1016/S0168-583X(00)00127-0} (\bibinfo{year}{2000}).

\bibitem{Zadvornaya2018}
\bibinfo{author}{Zadvornaya, A.} \emph{et~al.}
\newblock \bibinfo{journal}{\bibinfo{title}{{Characterization of Supersonic Gas
  Jets for High-Resolution Laser Ionization Spectroscopy of Heavy Elements}}}.
\newblock {\emph{\JournalTitle{Physical Review X}}}
  \textbf{\bibinfo{volume}{8}}, \bibinfo{pages}{041008},
  \doiprefix\url{10.1103/PhysRevX.8.041008} (\bibinfo{year}{2018}).

\bibitem{Jonsson1983}
\bibinfo{author}{J{\"{o}}nsson, G.}, \bibinfo{author}{Lundberg, H.} \&
  \bibinfo{author}{Svanberg, S.}
\newblock \bibinfo{journal}{\bibinfo{title}{{Lifetime measurements in the S1/2
  and D3/2, 5/2 sequences of indium}}}.
\newblock {\emph{\JournalTitle{Physical Review A}}}
  \textbf{\bibinfo{volume}{27}}, \bibinfo{pages}{2930--2935},
  \doiprefix\url{10.1103/PhysRevA.27.2930} (\bibinfo{year}{1983}).

\bibitem{Kessler2008}
\bibinfo{author}{Kessler, T.}, \bibinfo{author}{Tomita, H.},
  \bibinfo{author}{Mattolat, C.}, \bibinfo{author}{Raeder, S.} \&
  \bibinfo{author}{Wendt, K.}
\newblock \bibinfo{journal}{\bibinfo{title}{{An injection-seeded
  high-repetition rate Ti:Sapphire laser for high-resolution spectroscopy and
  trace analysis of rare isotopes}}}.
\newblock {\emph{\JournalTitle{Laser Physics}}} \textbf{\bibinfo{volume}{18}},
  \bibinfo{pages}{842--849}, \doiprefix\url{10.1134/S1054660X08070074}
  (\bibinfo{year}{2008}).

\bibitem{Sonnenschein2017}
\bibinfo{author}{Sonnenschein, V.} \emph{et~al.}
\newblock \bibinfo{journal}{\bibinfo{title}{{Characterization of a pulsed
  injection-locked Ti:sapphire laser and its application to high resolution
  resonance ionization spectroscopy of copper}}}.
\newblock {\emph{\JournalTitle{Laser Physics}}} \textbf{\bibinfo{volume}{27}},
  \bibinfo{pages}{085701}, \doiprefix\url{10.1088/1555-6611/aa7834}
  (\bibinfo{year}{2017}).

\bibitem{Sahoo2015}
\bibinfo{author}{Sahoo, B.~K.}, \bibinfo{author}{Nandy, D.~K.},
  \bibinfo{author}{Das, B.~P.} \& \bibinfo{author}{Sakemi, Y.}
\newblock \bibinfo{journal}{\bibinfo{title}{{Correlation trends in the
  hyperfine structures of 210 , 212 Fr}}}.
\newblock {\emph{\JournalTitle{Physical Review A}}}
  \textbf{\bibinfo{volume}{042507}}, \bibinfo{pages}{1--9},
  \doiprefix\url{10.1103/PhysRevA.91.042507} (\bibinfo{year}{2015}).

\bibitem{Yu2019}
\bibinfo{author}{Yu, Y.~M.} \& \bibinfo{author}{Sahoo, B.~K.}
\newblock \bibinfo{journal}{\bibinfo{title}{{Investigating ground-state
  fine-structure properties to explore suitability of boronlike
  S{\$}{\^{}}{\{}11+-{\}}{\$} K{\$}{\^{}}{\{}14+{\}}{\$} and galliumlike
  Nb{\$}{\^{}}{\{}10+-{\}}{\$} Ru{\$}{\^{}}{\{}13+{\}}{\$} ions as possible
  atomic clocks}}}.
\newblock {\emph{\JournalTitle{Physical Review A}}}
  \textbf{\bibinfo{volume}{99}}, \bibinfo{pages}{022513},
  \doiprefix\url{10.1103/PhysRevA.99.022513} (\bibinfo{year}{2019}).

\bibitem{Neijzen1982a}
\bibinfo{author}{Neijzen, J.~H.} \& \bibinfo{author}{D{\"{o}}nszelmann, A.}
\newblock \bibinfo{journal}{\bibinfo{title}{{Dye laser study of the np 2P1 2, 3
  2 Rydberg series in neutral gallium and indium atoms}}}.
\newblock {\emph{\JournalTitle{Physica B+C}}} \textbf{\bibinfo{volume}{114}},
  \bibinfo{pages}{241--250}, \doiprefix\url{10.1016/0378-4363(82)90043-2}
  (\bibinfo{year}{1982}).

\bibitem{NIST}
\bibinfo{author}{{A. Kramida, Y. Ralchenko, J. Reader}, N. A.~T.}
\newblock \bibinfo{title}{{NIST Atomic Spectra Database}}
  (\bibinfo{year}{2019}).

\bibitem{Hong1995}
\bibinfo{author}{Hong, F.~L.}, \bibinfo{author}{Maeda, H.},
  \bibinfo{author}{Matsuo, Y.} \& \bibinfo{author}{Takami, M.}
\newblock \bibinfo{journal}{\bibinfo{title}{{Inverted fine structure in highly
  excited 2F Rydberg states of indium}}}.
\newblock {\emph{\JournalTitle{Physical Review A}}}
  \textbf{\bibinfo{volume}{51}}, \bibinfo{pages}{1994--1998},
  \doiprefix\url{10.1103/PhysRevA.51.1994} (\bibinfo{year}{1995}).

\bibitem{Newville2014}
\bibinfo{author}{Newville, M.}, \bibinfo{author}{Stensitzki, T.},
  \bibinfo{author}{Allen, D.~B.} \& \bibinfo{author}{Ingargiola, A.}
\newblock \bibinfo{title}{{LMFIT: Non-Linear Least-Square Minimization and
  Curve-Fitting for Python}}, \doiprefix\url{10.5281/ZENODO.11813}
  (\bibinfo{year}{2014}).

\bibitem{Schwartz1955}
\bibinfo{author}{Schwartz, C.}
\newblock \bibinfo{journal}{\bibinfo{title}{{Theory of Hyperfine Structure}}}.
\newblock {\emph{\JournalTitle{Physical Review}}}
  \textbf{\bibinfo{volume}{97}}, \bibinfo{pages}{380--395},
  \doiprefix\url{10.1103/PhysRev.97.380} (\bibinfo{year}{1955}).

\bibitem{Olivero1977}
\bibinfo{author}{Olivero, J.} \& \bibinfo{author}{Longbothum, R.}
\newblock \bibinfo{journal}{\bibinfo{title}{{Empirical fits to the Voigt line
  width: A brief review}}}.
\newblock {\emph{\JournalTitle{Journal of Quantitative Spectroscopy and
  Radiative Transfer}}} \textbf{\bibinfo{volume}{17}},
  \bibinfo{pages}{233--236}, \doiprefix\url{10.1016/0022-4073(77)90161-3}
  (\bibinfo{year}{1977}).

\bibitem{Feneuille1982}
\bibinfo{author}{Feneuille, S.} \& \bibinfo{author}{Jacquinot, P.}
\newblock \bibinfo{journal}{\bibinfo{title}{{Atomic rydberg states}}}.
\newblock {\emph{\JournalTitle{Advances in Atomic and Molecular Physics}}}
  \textbf{\bibinfo{volume}{17}}, \bibinfo{pages}{99--166},
  \doiprefix\url{10.1016/S0065-2199(08)60068-8} (\bibinfo{year}{1982}).

\bibitem{Haynes}
\bibinfo{author}{Haynes, W.~M.}
\newblock \emph{\bibinfo{title}{{CRC handbook of chemistry and physics : a
  ready-reference book of chemical and physical data.}}}
  (\bibinfo{publisher}{CRC Press}, \bibinfo{address}{Boulder, Colorado, USA},
  \bibinfo{year}{2015}), \bibinfo{edition}{96th} edn.

\bibitem{GarciaRuiz2018a}
\bibinfo{author}{{Garcia Ruiz}, R.~F.} \emph{et~al.}
\newblock \bibinfo{journal}{\bibinfo{title}{{High-Precision Multiphoton
  Ionization of Accelerated Laser-Ablated Species}}}.
\newblock {\emph{\JournalTitle{Physical Review X}}}
  \textbf{\bibinfo{volume}{8}}, \bibinfo{pages}{041005},
  \doiprefix\url{10.1103/PhysRevX.8.041005} (\bibinfo{year}{2018}).

\bibitem{Fricke}
\bibinfo{author}{Fricke, G.} \& \bibinfo{author}{Heilig, K.}
\newblock \bibinfo{title}{{49-In Indium}}.
\newblock In \emph{\bibinfo{booktitle}{Nuclear Charge Radii}},
  \bibinfo{pages}{1--6}, \doiprefix\url{10.1007/10856314_51}
  (\bibinfo{publisher}{Springer-Verlag}, \bibinfo{address}{Berlin/Heidelberg},
  \bibinfo{year}{2004}).

\bibitem{Rice1957}
\bibinfo{author}{Rice, M.} \& \bibinfo{author}{Pound, R.~V.}
\newblock \bibinfo{journal}{\bibinfo{title}{{Ratio of the Magnetic Moments of
  In 115 and In 113}}}.
\newblock {\emph{\JournalTitle{Physical Review}}}
  \textbf{\bibinfo{volume}{106}}, \bibinfo{pages}{953--953},
  \doiprefix\url{10.1103/PhysRev.106.953} (\bibinfo{year}{1957}).

\bibitem{Flynn1960}
\bibinfo{author}{Flynn, C.~P.} \& \bibinfo{author}{Seymour, E. F.~W.}
\newblock \bibinfo{journal}{\bibinfo{title}{{Knight Shift of the Nuclear
  Magnetic Resonance in Liquid Indium}}}.
\newblock {\emph{\JournalTitle{Proceedings of the Physical Society}}}
  \textbf{\bibinfo{volume}{76}}, \bibinfo{pages}{301--303},
  \doiprefix\url{10.1088/0370-1328/76/2/415} (\bibinfo{year}{1960}).

\bibitem{Wang2017}
\bibinfo{author}{Wang, M.} \emph{et~al.}
\newblock \bibinfo{journal}{\bibinfo{title}{{The AME2016 atomic mass evaluation
  (II). Tables, graphs and references}}}.
\newblock {\emph{\JournalTitle{Chinese Physics C}}}
  \textbf{\bibinfo{volume}{41}}, \bibinfo{pages}{30003},
  \doiprefix\url{10.1088/1674-1137/41/3/030003} (\bibinfo{year}{2017}).

\bibitem{George1990b}
\bibinfo{author}{George, S.}, \bibinfo{author}{Verges, J.} \&
  \bibinfo{author}{Guppy, G.}
\newblock \bibinfo{journal}{\bibinfo{title}{{Newly observed lines and hyperfine
  structure in the infrared spectrum of indium obtained by using a
  Fourier-transform spectrometer}}}.
\newblock {\emph{\JournalTitle{Journal of the Optical Society of America B}}}
  \textbf{\bibinfo{volume}{7}}, \bibinfo{pages}{249},
  \doiprefix\url{10.1364/josab.7.000249} (\bibinfo{year}{1990}).

\bibitem{Koszorus2019a}
\bibinfo{author}{Koszor{\'{u}}s} \emph{et~al.}
\newblock \bibinfo{journal}{\bibinfo{title}{{Precision measurements of the
  charge radii of potassium isotopes}}}.
\newblock {\emph{\JournalTitle{Physical Review C}}}
  \textbf{\bibinfo{volume}{100}}, \doiprefix\url{10.1103/PhysRevC.100.034304}
  (\bibinfo{year}{2019}).

\bibitem{Neijzen1981}
\bibinfo{author}{Neijzen, J.~H.} \& \bibinfo{author}{D{\"{o}}nszelmann, A.}
\newblock \bibinfo{journal}{\bibinfo{title}{{A study of the np 2P1 2, 3 2
  Rydberg series in neutral indium by means of two-photon laser
  spectroscopy}}}.
\newblock {\emph{\JournalTitle{Physica B+C}}} \textbf{\bibinfo{volume}{111}},
  \bibinfo{pages}{127--133}, \doiprefix\url{10.1016/0378-4363(81)90171-6}
  (\bibinfo{year}{1981}).

\bibitem{Rothe2013}
\bibinfo{author}{Rothe, S.} \emph{et~al.}
\newblock \bibinfo{journal}{\bibinfo{title}{{Measurement of the first
  ionization potential of astatine by laser ionization spectroscopy}}}.
\newblock {\emph{\JournalTitle{Nature Communications}}}
  \textbf{\bibinfo{volume}{4}}, \doiprefix\url{10.1038/ncomms2819}
  (\bibinfo{year}{2013}).

\bibitem{Foot2004}
\bibinfo{author}{Foot, C.}
\newblock \emph{\bibinfo{title}{{Atomic Physics}}}, vol.~\bibinfo{volume}{25}
  (\bibinfo{publisher}{OUP Oxford}, \bibinfo{year}{2004}).

\bibitem{Drake1994}
\bibinfo{author}{Drake, G.~W.}
\newblock \bibinfo{journal}{\bibinfo{title}{{Quantum Defect Theory and Analysis
  of High-Precision Helium Term Energies}}}.
\newblock {\emph{\JournalTitle{Advances in Atomic, Molecular and Optical
  Physics}}} \textbf{\bibinfo{volume}{32}}, \bibinfo{pages}{93--116},
  \doiprefix\url{10.1016/S1049-250X(08)60012-9} (\bibinfo{year}{1994}).

\bibitem{Mount2009}
\bibinfo{author}{Mount, B.~J.}, \bibinfo{author}{Redshaw, M.} \&
  \bibinfo{author}{Myers, E.~G.}
\newblock \bibinfo{journal}{\bibinfo{title}{{Q Value of 115In to 115Sn (3/2+):
  The Lowest Known Energy $\beta$ Decay}}}.
\newblock {\emph{\JournalTitle{Physical Review Letters}}}
  \textbf{\bibinfo{volume}{103}},
  \doiprefix\url{10.1103/PhysRevLett.103.122502} (\bibinfo{year}{2009}).

\bibitem{Wieslander2009}
\bibinfo{author}{Wieslander, J.~S.} \emph{et~al.}
\newblock \bibinfo{journal}{\bibinfo{title}{{Smallest Known Q Value of Any
  Nuclear Decay: The Rare beta- Decay of In115(9/2+) to Sn115(3/2+)}}}.
\newblock {\emph{\JournalTitle{Physical Review Letters}}}
  \textbf{\bibinfo{volume}{103}}, \bibinfo{pages}{122501},
  \doiprefix\url{10.1103/PhysRevLett.103.122501} (\bibinfo{year}{2009}).

\bibitem{Neijzen1981a}
\bibinfo{author}{Neijzen, J.~H.} \& \bibinfo{author}{D{\"{o}}nszelmann, A.}
\newblock \bibinfo{journal}{\bibinfo{title}{{Configuration interaction effects
  in the 2S1/2, 2D3/2,5/2 rydberg series of neutral indium investigated with a
  frequency-doubled dye laser}}}.
\newblock {\emph{\JournalTitle{Physica B+C}}} \textbf{\bibinfo{volume}{106}},
  \bibinfo{pages}{271--286}, \doiprefix\url{10.1016/0378-4363(81)90087-5}
  (\bibinfo{year}{1981}).

\bibitem{Martin1980}
\bibinfo{author}{Martin, W.~C.}
\newblock \bibinfo{journal}{\bibinfo{title}{{Series formulas for the spectrum
  of atomic sodium (Na I)}}}.
\newblock {\emph{\JournalTitle{Journal of the Optical Society of America}}}
  \textbf{\bibinfo{volume}{70}}, \bibinfo{pages}{784},
  \doiprefix\url{10.1364/josa.70.000784} (\bibinfo{year}{1980}).

\bibitem{George1990c}
\bibinfo{author}{George, S.}, \bibinfo{author}{Verges, J.} \&
  \bibinfo{author}{Guppy, G.}
\newblock \bibinfo{journal}{\bibinfo{title}{{Newly observed lines and hyperfine
  structure in the infrared spectrum of indium obtained by using a
  Fourier-transform spectrometer}}}.
\newblock {\emph{\JournalTitle{Journal of the Optical Society of America B}}}
  \textbf{\bibinfo{volume}{7}}, \bibinfo{pages}{249},
  \doiprefix\url{10.1364/josab.7.000249} (\bibinfo{year}{1990}).

\bibitem{Johansson1967}
\bibinfo{author}{Johansson, I.} \& \bibinfo{author}{Litzen, U.}
\newblock \bibinfo{journal}{\bibinfo{title}{{Term systems of neutral gallium
  and indium atoms derived from new measurements in infared region}}}.
\newblock {\emph{\JournalTitle{ARKIV FOR FYSIK}}}
  \textbf{\bibinfo{volume}{34}}, \bibinfo{pages}{573} (\bibinfo{year}{1967}).

\bibitem{moore1949atomic}
\bibinfo{author}{Moore, C.~E.}
\newblock \emph{\bibinfo{title}{{Atomic Energy Levels as Derived from the
  Analyses of Optical Spectra: The spectra of hydrogen, deuterium, tritium,
  helium, lithium, beryllium, boron, carbon, nitrogen, oxygen, flourine, neon,
  sodium, magnesium, aluminum, silicon, phosphorus, sulfur, chlor}}}.
\newblock Atomic Energy Levels as Derived from the Analyses of Optical Spectra
  (\bibinfo{publisher}{U.S. Department of Commerce, National Bureau of
  Standards}, \bibinfo{year}{1949}).

\bibitem{Das2011}
\bibinfo{author}{Das, M.}, \bibinfo{author}{Chaudhuri, R.~K.},
  \bibinfo{author}{Chattopadhyay, S.} \& \bibinfo{author}{Mahapatra, U.~S.}
\newblock \bibinfo{journal}{\bibinfo{title}{{Valence universal multireference
  coupled cluster calculations of the properties of indium in its ground and
  excited states}}}.
\newblock {\emph{\JournalTitle{Journal of Physics B: Atomic, Molecular and
  Optical Physics}}} \textbf{\bibinfo{volume}{44}}, \bibinfo{pages}{065003},
  \doiprefix\url{10.1088/0953-4075/44/6/065003} (\bibinfo{year}{2011}).

\bibitem{Stratmann1994a}
\bibinfo{author}{Stratmann, K.} \emph{et~al.}
\newblock \bibinfo{journal}{\bibinfo{title}{{High-resolution field ionizer for
  state-selective detection of Rydberg atoms in fast-beam laser
  spectroscopy}}}.
\newblock {\emph{\JournalTitle{Review of Scientific Instruments}}}
  \textbf{\bibinfo{volume}{65}}, \bibinfo{pages}{1847--1852},
  \doiprefix\url{10.1063/1.1144833} (\bibinfo{year}{1994}).

\bibitem{AB}
\bibinfo{author}{COMSOL, A.~B.}
\newblock \bibinfo{title}{{COMSOL Multiphysics}}.

\bibitem{Robicheaux1997}
\bibinfo{author}{Robicheaux, F.}
\newblock \bibinfo{journal}{\bibinfo{title}{{Pulsed field ionization of Rydberg
  atoms}}}.
\newblock {\emph{\JournalTitle{Physical Review A - Atomic, Molecular, and
  Optical Physics}}} \textbf{\bibinfo{volume}{56}},
  \bibinfo{pages}{R3358--R3361}, \doiprefix\url{10.1103/PhysRevA.56.R3358}
  (\bibinfo{year}{1997}).

\bibitem{Baranov1994}
\bibinfo{author}{Baranov, L.~Y.}, \bibinfo{author}{Kris, R.},
  \bibinfo{author}{Levine, R.~D.}, \bibinfo{author}{Even, U.} \&
  \bibinfo{author}{Baranov, L.~V.}
\newblock \bibinfo{journal}{\bibinfo{title}{{On the field ionization spectrum
  of high Rydberg states}}}.
\newblock {\emph{\JournalTitle{The Journal of Chemical Physics}}}
  \textbf{\bibinfo{volume}{100}}, \bibinfo{pages}{3495},
  \doiprefix\url{10.1063/1.466978} (\bibinfo{year}{1994}).

\bibitem{Jones1993}
\bibinfo{author}{Jones, R.~R.}, \bibinfo{author}{You, D.} \&
  \bibinfo{author}{Bucksbaum, P.~H.}
\newblock \bibinfo{journal}{\bibinfo{title}{{Ionization of Rydberg atoms by
  subpicosecond half-cycle electromagnetic pulses}}}.
\newblock {\emph{\JournalTitle{Physical Review Letters}}}
  \textbf{\bibinfo{volume}{70}}, \bibinfo{pages}{1236--1239},
  \doiprefix\url{10.1103/PhysRevLett.70.1236} (\bibinfo{year}{1993}).

\bibitem{Ducas1975}
\bibinfo{author}{Ducas, T.~W.}, \bibinfo{author}{Littman, M.~G.},
  \bibinfo{author}{Freeman, R.~R.} \& \bibinfo{author}{Kleppner, D.}
\newblock \bibinfo{journal}{\bibinfo{title}{{Stark Ionization of High-Lying
  States of Sodium}}}.
\newblock {\emph{\JournalTitle{Physical Review Letters}}}
  \textbf{\bibinfo{volume}{35}}, \bibinfo{pages}{366--369},
  \doiprefix\url{10.1103/PhysRevLett.35.366} (\bibinfo{year}{1975}).

\bibitem{Siegman1986}
\bibinfo{author}{Siegman, A.~E.}
\newblock \emph{\bibinfo{title}{{Lasers}}} (\bibinfo{publisher}{University
  Science Books}, \bibinfo{address}{Mill Valley Calif.}, \bibinfo{year}{1986}).

\bibitem{Wang2014}
\bibinfo{author}{Wang, X.} \emph{et~al.}
\newblock \bibinfo{journal}{\bibinfo{title}{{Ion kinetic energy distributions
  in laser-induced plasma}}}.
\newblock {\emph{\JournalTitle{Spectrochimica Acta Part B: Atomic
  Spectroscopy}}} \textbf{\bibinfo{volume}{99}}, \bibinfo{pages}{101--114},
  \doiprefix\url{10.1016/J.SAB.2014.06.018} (\bibinfo{year}{2014}).

\bibitem{Burnett1993}
\bibinfo{author}{Burnett, K.}, \bibinfo{author}{Reed, V.~C.} \&
  \bibinfo{author}{Knight, P.~L.}
\newblock \bibinfo{journal}{\bibinfo{title}{{Atoms in ultra-intense laser
  fields}}}.
\newblock {\emph{\JournalTitle{Journal of Physics B: Atomic, Molecular and
  Optical Physics}}} \textbf{\bibinfo{volume}{26}}, \bibinfo{pages}{561--598},
  \doiprefix\url{10.1088/0953-4075/26/4/003} (\bibinfo{year}{1993}).

\bibitem{Shahzada2012}
\bibinfo{author}{Shahzada, S.} \emph{et~al.}
\newblock \bibinfo{journal}{\bibinfo{title}{{Photoionization studies from the
  3p P2 excited state of neutral lithium}}}.
\newblock {\emph{\JournalTitle{Journal of the Optical Society of America B}}}
  \textbf{\bibinfo{volume}{29}}, \bibinfo{pages}{3386},
  \doiprefix\url{10.1364/josab.29.003386} (\bibinfo{year}{2012}).

\bibitem{Kunc1991}
\bibinfo{author}{Kunc, J.~a.} \& \bibinfo{author}{Soon, W.~H.}
\newblock \bibinfo{journal}{\bibinfo{title}{{Analytical ionization cross
  sections for atomic collisions}}}.
\newblock {\emph{\JournalTitle{The Journal of Chemical Physics}}}
  \textbf{\bibinfo{volume}{95}}, \bibinfo{pages}{5738},
  \doiprefix\url{10.1063/1.461622} (\bibinfo{year}{1991}).

\bibitem{Cooper2019}
\bibinfo{author}{Cooper, B.~S.} \emph{et~al.}
\newblock \bibinfo{journal}{\bibinfo{title}{{A compact RFQ cooler buncher for
  CRIS experiments}}}.
\newblock {\emph{\JournalTitle{Hyperfine Interactions}}}
  \textbf{\bibinfo{volume}{240}}, \bibinfo{pages}{1--8},
  \doiprefix\url{10.1007/s10751-019-1586-7} (\bibinfo{year}{2019}).

\bibitem{Harasimowicz2012}
\bibinfo{author}{Harasimowicz, J.}, \bibinfo{author}{Welsch, C.~P.},
  \bibinfo{author}{Cosentino, L.}, \bibinfo{author}{Pappalardo, A.} \&
  \bibinfo{author}{Finocchiaro, P.}
\newblock \bibinfo{journal}{\bibinfo{title}{{Beam diagnostics for low energy
  beams}}}.
\newblock {\emph{\JournalTitle{Physical Review Special Topics - Accelerators
  and Beams}}} \textbf{\bibinfo{volume}{15}}, \bibinfo{pages}{122801},
  \doiprefix\url{10.1103/PhysRevSTAB.15.122801} (\bibinfo{year}{2012}).

\bibitem{Thompson1976}
\bibinfo{author}{Thompson, W.} \& \bibinfo{author}{Hanrahan, S.}
\newblock \bibinfo{journal}{\bibinfo{title}{{CHARACTERISTICS OF A CRYOGENIC
  EXTREME HIGH-VACUUM CHAMBER.}}}
\newblock {\emph{\JournalTitle{J Vac Sci Technol}}}
  \textbf{\bibinfo{volume}{14}}, \bibinfo{pages}{643--645},
  \doiprefix\url{10.1116/1.569168} (\bibinfo{year}{1976}).

\bibitem{Stroberg2019a}
\bibinfo{author}{Stroberg, S.~R.}, \bibinfo{author}{Bogner, S.~K.},
  \bibinfo{author}{Hergert, H.} \& \bibinfo{author}{Holt, J.~D.}
\newblock \bibinfo{journal}{\bibinfo{title}{{Nonempirical Interactions for the
  Nuclear Shell Model: An Update}}}.
\newblock {\emph{\JournalTitle{Annual Review of Nuclear and Particle Science}}}
  \textbf{\bibinfo{volume}{69}},
  \doiprefix\url{10.1146/annurev-nucl-101917-021120} (\bibinfo{year}{2019}).
\newblock \eprint{1902.06154}.

\bibitem{Ekstrom2019a}
\bibinfo{author}{Ekstr{\"{o}}m, A.} \& \bibinfo{author}{Hagen, G.}
\newblock \bibinfo{journal}{\bibinfo{title}{{Global Sensitivity Analysis of
  Bulk Properties of an Atomic Nucleus}}}.
\newblock {\emph{\JournalTitle{Physical Review Letters}}}
  \textbf{\bibinfo{volume}{123}}, \bibinfo{pages}{252501},
  \doiprefix\url{10.1103/PhysRevLett.123.252501} (\bibinfo{year}{2019}).
\newblock \eprint{1910.02922}.

\bibitem{Morris2018}
\bibinfo{author}{Morris, T.~D.} \emph{et~al.}
\newblock \bibinfo{journal}{\bibinfo{title}{{Structure of the Lightest Tin
  Isotopes}}}.
\newblock {\emph{\JournalTitle{Physical Review Letters}}}
  \textbf{\bibinfo{volume}{120}}, \bibinfo{pages}{152503},
  \doiprefix\url{10.1103/PhysRevLett.120.152503} (\bibinfo{year}{2018}).

\bibitem{Carlson2015b}
\bibinfo{author}{Carlson, J.} \emph{et~al.}
\newblock \bibinfo{journal}{\bibinfo{title}{{Quantum Monte Carlo methods for
  nuclear physics}}}.
\newblock {\emph{\JournalTitle{Reviews of Modern Physics}}}
  \textbf{\bibinfo{volume}{87}}, \doiprefix\url{10.1103/RevModPhys.87.1067}
  (\bibinfo{year}{2015}).
\newblock \eprint{1412.3081}.

\end{thebibliography}

\end{document}